\documentclass{emulateapj}
\usepackage{amsmath}
\let\oldAA\AA
\renewcommand{\AA}{\text{\normalfont\oldAA}}
\usepackage{setspace} 
\usepackage{graphicx}
\usepackage[english]{babel}
\usepackage[utf8]{inputenc} 
\usepackage{slashed}
\usepackage[backref,breaklinks,colorlinks,citecolor=blue]{hyperref}
\usepackage{epstopdf}
\usepackage{natbib}
\usepackage{amsmath} 
\newcommand{\gt}{>}
\newcommand{\civ}{\ifmmode {\rm C}\,\textsc{iv}  \else C\,\textsc{iv}\fi}
\newcommand{\CIV}{\ifmmode {\rm C}\,\textsc{iv}\,\lambda1549 \else C\,\textsc{iv}\,$\lambda1549$\fi}
\newcommand{\mgii}{\ifmmode {\rm Mg}\,\textsc{ii} \else Mg\,\textsc{ii}\fi}
\newcommand{\MgII}{\ifmmode {\rm Mg}\,\textsc{ii}\,\lambda2798 \else Mg\,\textsc{ii}\,$\lambda2798$\fi}
\begin{document}

\title{Optical variability of AGN in the PTF/iPTF survey}
\author{Neven Caplar}
 \email{neven.caplar@phys.ethz.ch}
\author{Simon J. Lilly}
\author{Benny Trakhtenbrot\altaffilmark{1}}
\altaffiltext{1}{Zwicky Fellow}
\affiliation{%
Institute for Astronomy, Department of Physics, ETH Zurich, Wolfgang-Pauli-Strasse 27, 8093, Zurich, Switzerland
}%
  \today

\begin{abstract}

We characterize the optical variability of quasars in the intermediate Palomar Transient Factory (iPTF) and Palomar Transient Factory (PTF) surveys. 
We re-calibrate the $r$-band light curves for $\sim$28,000 luminous, broad-line AGNs from the SDSS, producing a total of $\sim$2.4 million photometric data points. 
We utilize both the structure function (SF) and power spectrum density (PSD) formalisms to search for links between the optical variability and the physical parameters of the accreting supermassive black holes that power the quasars. 
The excess variance (SF$^{2}$) of the quasar sample tends to zero at very short time separations, validating our re-calibration of the time-series data. 
We find that the the amplitude of variability at a given time-interval, or equivalently the time-scale of variability to reach a certain amplitude, is most strongly correlated with luminosity with weak or no dependence on black hole mass and redshift. 
For a variability level of SF($\tau$)=0.07 mag, the time-scale has a dependency of $\tau \propto L^{0.4}$.
This is broadly consistent with the expectation from a simple Keplerian accretion disk model, which provides $\tau \propto L^{0.5}$. 
The PSD analysis also reveals that many quasar light curves are steeper than a damped random walk. We find a correlation between the steepness of the PSD slopes, specifically the fraction of slopes steeper than 2.5, and black hole mass, although we cannot exclude the possibility that luminosity or Eddington ratio are the drivers of this effect.  This effect is also seen in the SF analysis of the (i)PTF data, and in a PSD analysis of quasars in the SDSS Stripe 82.

\end{abstract}
\maketitle

\section{Introduction}

Time variability is potentially one of the most interesting properties of quasars and active galactic nuclei (AGN) generally. The variability of quasars and AGN (we will henceforth use these terms interchangeably) was recognized soon after the discovery of these objects \citep{Mat63}. It is however still largely unexplained and remains an active area of research. Variability in the brightness of AGN has been directly observed over a very wide range of time-scales, ranging from minutes and hours (e.g. \citealp{Kas15}, \citealp{McH16}) up to several years \citep{Mac10} and even, by combining data from optical plates with modern surveys, to over 50 years (\citealp{deV03}, \citealp{Ses06}). Indirect observational evidence exists for variability on scales of $\sim 10^{5}$ years \citep{Sch15} while simulations predict large scale variability during the AGN phase that may last $\sim 10^7$ years \citep{Nov11}.  \\

Quasar variability has been extensively studied in the optical waveband. The first study with very large numbers of objects used multiply-imaged regions of the SDSS survey \citep{Van04}. Further progress has been achieved by dedicated surveys of the variable sky, such as OGLE \citep{K10}, the Palomar-QUEST Survey \citep{Bau09}, the SDSS Stripe 82 (\citealp{Ses07}, \citealp{Mac10}, \citealp{Mac12}), PanStarrs (\citealp{Mor14}, \citealp{Sim16}), SUDARE-VOICE \citep{F15} and LaSilla-Quest \citep{Car15}. \\
	
These studies have in general found an anti-correlation of the amplitude of variability (at a fixed rest-frame timescale) with luminosity, an increase of variability towards shorter wavelengths and little to no apparent dependence on redshift.   The relationship with black hole mass remains unclear, with different studies finding either positive, negative or absent correlations (\citealp{Wol07}, \citealp{Wil08}, \citealp{Kel09}, \citealp{Mac10}, \citealp{Zuo12}).   In the vast majority of cases \citep{And13}, the variability is found to be of stochastic nature, with only a few recent claims of periodicity in a small number of objects (\citealp{Gra15}, \citealp{Gra152}, \citealp{Cha16}, but see also \citealp{Vau16}). To describe the characteristics of this stochastic variability, the ``damped random walk'' model, first suggested in \cite{Kel09}, is usually used.   On short time scales, the variability is dominated by short-term random changes, producing a red noise spectrum.  On longer timescales, the variation of brightness becomes uncorrelated, leading to a flattening of the spectrum towards white noise.  The Kepler mission, with its precise photometry and continuous dense sampling of the optical light curves, has provided new insights and produced unambiguous power spectral density (PSD) curves for a small number of AGN. These individual PSD are often found to be inconsistent with the damped-random walk model, especially on short timescales  \citep{Kas15}.    \\

A range of possible explanations for brightness variations in quasars have been proposed.  These include micro-lensing, accretion disk instabilities and variations in the accretion rate (\citealp{Haw93}, \citealp{Are97}, \citealp{Kaw98},\citealp{Tre02}, \citealp{Haw07}).   The fact that the observed variability in the optical bands can be seen on scales of days to weeks is hard to reconcile with a model in which variability is caused by changes in the accretion rate. This is because the associated viscous time scales that should describe the radial migration or ``drift" through the disk would be of order $10^4$ years for typical AGN \citep{Net13}. On the other hand accretion disk instabilities would be connected with the dynamical or orbital time scales, which are much shorter and will be of the order of $\sim 1$ yr for typical AGN. Thin accretion disk theory also predicts the dependence of these time scales with other physical parameters. Under the simplest assumptions. the viscous time scale varies as $\tau_{\rm vis} \propto  L^{7/60} \lambda^{5/3}_{RF}M^{2/3}_{bh}$ , while the dynamical time scale should go as $\tau_{\rm dyn} \propto L^{1/2} \lambda^{2}_{RF}$ \citep{Mac10}. 
Models with independent emitting regions that experience localized temperature variations, possibly created by instabilities in the accretion flow, are able to explain, in a qualitative manner, a wide range of features seen in optical variability surveys.  They are also able to explain the measurements of large sizes of AGN disks from micro-lensing (\citealp{Dex11}, \citealp{Cai16}).  \\

Ground-based optical surveys generally suffer from irregular sampling, because of the realities of nightly, monthly and yearly observing periods. Therefore it is necessary to use alternative methods beyond Fourier analysis to characterize the variability. The most commonly used approach is to construct the structure function (SF).  The SF is defined to be the rms magnitude difference (in excess of observational noise) as a function of the time difference $\Delta t$ between pairs of measurements. In this paper we will generally use the SF$^{2}$ which is the excess variance (arising from variability) of the magnitude difference as a function of $\Delta t$:
\begin{equation} \label{eq: SF}
\mbox{SF}^{2} (\Delta t) = \frac{1}{P} \sum^{P}_{i, j>i} \left( m_{i}-m_{j} \right)^{2}-\sigma^{2}_{i}-\sigma^{2}_{j}  
\end{equation}
 where the sum is over all $P$ measurements which are separated by some time lag $\Delta t$, taking into account the measurement errors $\sigma$ of the individual data points. The recent study of \cite{Koz16} offers a good overview of the SF formalism, discusses different definitions used in the literature and points out a number of caveats. Using the SF formalism, it is possible to construct estimates of the ensemble variability of a set of objects.  This implicitly assumes that an analysis of the variability of the members of a suitably defined set of quasars will reveal the same physical relations as studying individual objects.  \\

AGNs have been found to vary across all observational wavebands and PSD analysis has proved useful in the X-ray band because the time sampling is generally regular.  The PSDs are in general found to be well described by a broken power law with a slope of roughly $ \nu^{-2}$ (as in a random walk) above a break frequency $\nu_{br}$ and a flatter slope below the break frequency (\citealp{Law93}, \citealp{Gre93}), broadly mimicking the behaviour of the SF at optical wavelengths.  The exact nature of the phenomenological correlation found between the break frequency and the mass of the black hole ($m_{bh}$) is still unclear (\citealp{Mch06}, \citealp{Gon12}, \citealp{Emm16}).  \\

At optical wavelengths, the PSD formalism has been used in the case of the regularly and densely sampled Kepler data.   These data suggest a wide spread of values for the high frequency slope of the optical PSD, with indicated slopes of between 2 and 4 (\citealp{Mus11}, \citealp{Ede14}). The existence of a break frequency and associated flattening of the slope at low frequencies is still unclear, with various studies suggesting breaks occuring at scales of anywhere from $\sim$10 to $\sim$ 1000 days or longer (\citealp{Col01}, \citealp{Kel09}, \citealp{Mac10}).  Recently \cite{Kel14} introduced a flexible algorithm to estimate the PSD of light curves in the context of a broad family of continuous-time autoregressive moving average processes, which also enables estimates of PSD for sparsely sampled data to be obtained.  \\

In this work we aim to characterise the optical variability of a very large number of quasars in the Palomar Transient Factory (PTF) and intermediate Palomar Transient Factory (iPTF) surveys.  The analysis involves a complete photometric recalibration of the photometry of the quasars in these surveys. The combination of a very large number of photometric data points and the wide sky coverage of PTF and iPTF enables us to study AGN variability within what is believed to be the largest fully-calibrated single-band dataset available.  We then investigate the connections between the optical variability and the physical properties of the AGNs in terms of luminosity, black hole mass and redshift.   In general, we use time domain analysis, using the SF formalism, as well as new methods introduced in \cite{Kel14} to study variability in the frequency domain. \\

The paper is organized as follows. In Section 2 we describe the quasar sample. In Section 3 we describe the re-calibration of the photometry while in Section 4 we explain the methods used in the SF and PSD analyses.  
We then focus on two aspects of variability. In Section 5 we explore the correlations between physical parameters of the quasars, specifically luminosity, black hole mass and redshift, and the amplitude of variability (on a given time-scale) as well as the largely equivalent time-scale of variability at a given amplitude, as derived from the SF analysis.  We clearly see a number of results that have been indicated previously in generally smaller data sets.   The strongest correlations are seen with quasar luminosity with weak or absent trends with black hole mass and redshift.    In Section 6 we explore the correlations of the slopes of the PSD with the physical parameters, arguing that quasars with more massive black holes exhibit steeper PSD. We show that this correlation can be seen in SDSS data, and is also consistent with the previous SF analysis. We further discuss how to connect the result from the SF and the PSD analysis in Section 7. We summarise our conclusions in Section 8. In the Appendices we expand on several topics concerning technical aspects of the re-calibration.   \\

\section{The sample \& data}

The PTF/iPTF is studying the transient sky using the 48$^{''}$ telescope on Mount Palomar.  It employs the former Canada-France-Hawaii Telescope 12K $\times$ 8K camera with 11 active chips which together cover an effective 7.26 deg$^{2}$ field of view (\citealp{Rah08},  \citealp{Law09}, \citealp{Rau09}). Most of the PTF/iPTF observations, which started in the March 2009, are taken in the Mould $\left\langle r \right\rangle $ filter ($\lambda$ = 6580 $\AA$). The median seeing of the images is about  2.2$^{''}$. Individual exposures are 60 seconds and the surveys cover $\sim$ \mbox{8000 deg$^{2}$} of the northern sky above declination of $-30^{\circ}$. Further information about the performance of the survey and the data preparation procedures are available in \cite{Law09} and \cite{Lah14}. Initial calibration for the survey was reported in \cite{Ofe12}, but we decided to recalibrate the quasar data with our specific scientific goals in mind. This re-calibration is presented below in Section \ref{sec:Cal}.\\

\begin{figure}[htp]
    \centering
    \includegraphics[width=.49\textwidth]{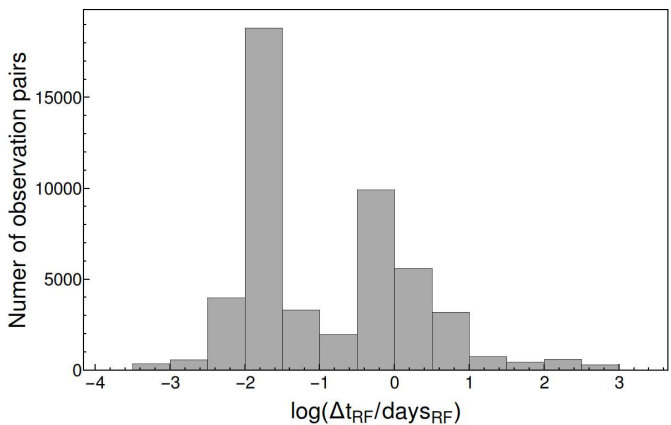}
    \caption{Distribution of observational cadence in the sample. The time differences have been transformed into the quasar restframe.}
\end{figure}

\begin{figure}[htp]
    \centering
   \includegraphics[width=.49\textwidth]{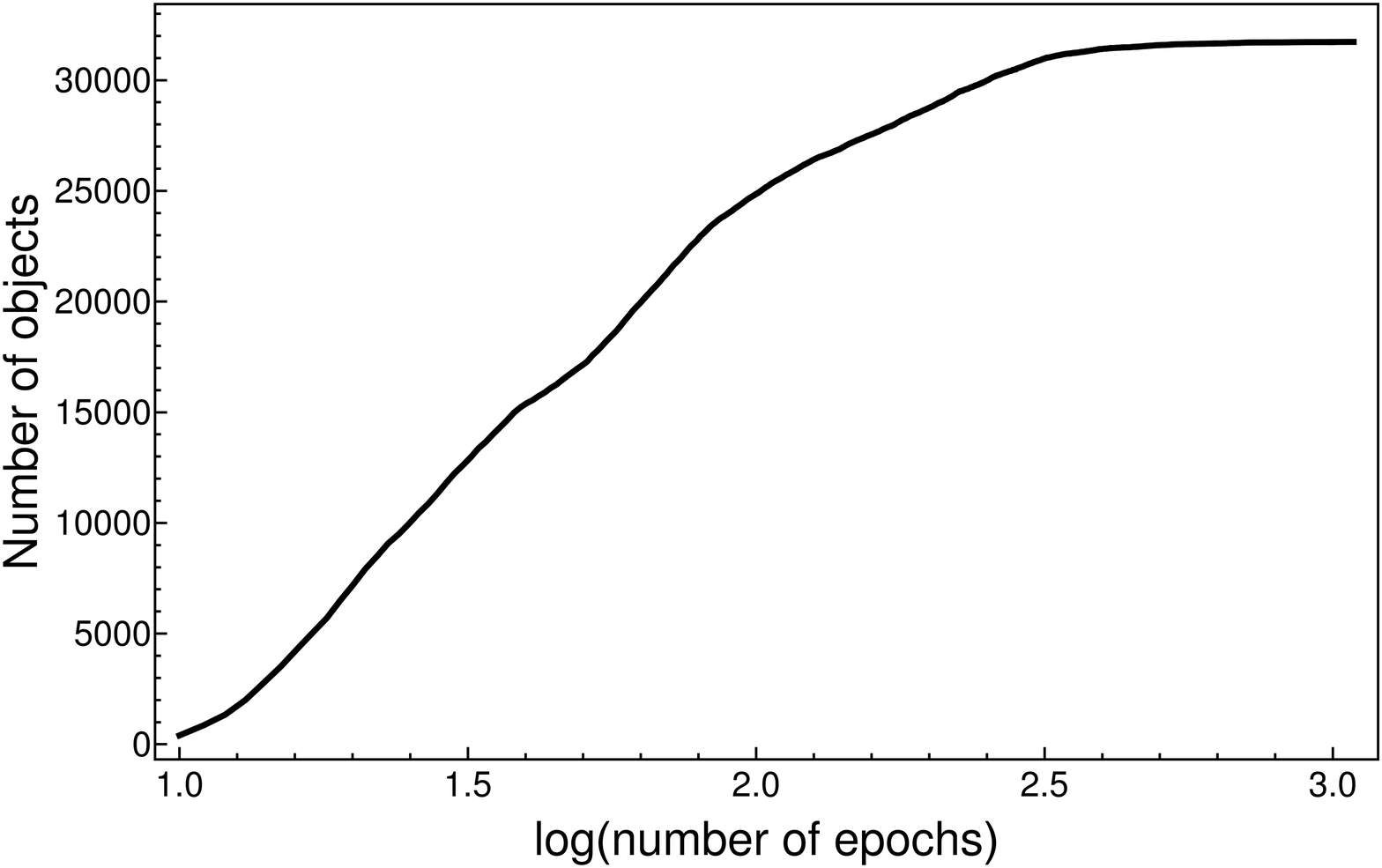}
    \caption{Cumulative distribution of the number of observations for the objects in the sample. There are $\sim$ \textbf{14,500} quasars with light curves with more than 50 observations, and $\sim$\textbf{1,750} light-curves with more than 250 observations.}
\end{figure}

\begin{figure}[htp]
    \centering
    \includegraphics[width=.49\textwidth]{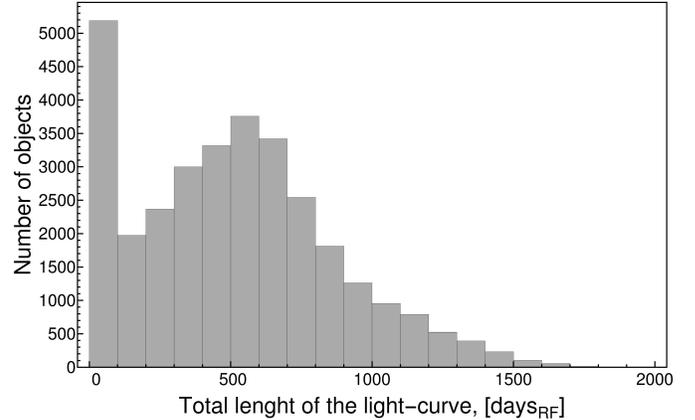}
    \caption{Cumulative distribution of the time span of the observations for the objects in the sample, i.e. the total length of the light curve between the first and the last observation. The time spans have been transformed into the quasar restframe.}
\end{figure}

This analysis is based on all the observations that were taken in the $r$-band in the six year interval between the start of the survey in March 2009 and April 2015. A small fraction of images were taken with $g$-band and H$\alpha$ filters, but these were not included due to the low number of available observations. We only use images which are not significantly affected by clouds, i.e. if the mean observed instrumental magnitudes of our reference objects (see below) are within 0.2 mag of the instrumental magnitudes in the 5 best observations of that field. This cut  removes a significant fraction of the available data ($\sim$30\%), but is necessary in order to achieve a high quality of calibration (see also Appendix A), not least because, with such a large field of view and short exposure, the effects of clouds can vary significantly across the field of view.  The survey consists of a large number of images of specific regions of the sky, or ``fields'', each-defined by a single field-center. Our work only includes those fields for which images at more than 10 epochs are available.  \\

The sample of quasars was constructed from the homogeneous sample of AGN identified in the Sloan Digital Sky Survey Data Release 7 (SDSS-DR7) by \cite{She11}. All physical quantities are taken from the accompanying catalog, except for black hole masses which are taken from a catalog with improved estimates in \cite{Tra12}. In cases where mass estimates from \cite{Tra12} are not available, we retain the \cite{She11} mass estimates.  We have verified that the choice of the black hole mass catalog makes no significant difference to our final conclusions. 
In both of the catalogs, black hole masses are estimated using the single-epoch method in which the mass is estimated from the width of the broad emission lines (H$\beta$, \MgII\hspace{0.15 cm}or \CIV) combined with the monochromatic luminosity at 5100, 3000 or  1450$\AA$, respectively. The exact choice of the line and monochromatic luminosity that is used depends on the redshift of the source. \\

We have then selected only quasars for which the median instrumental magnitude in the PTF/iPTF was brighter than -9 mag which corresponds broadly to $r=19.1$ mag.\footnote{Instrumental magnitudes refer to the measured brightness of the sources, as obtained from the automatically generated catalogs provided by the PTF collaboration. Details regarding these measurements are given in \cite{Lah14}.} 
We impose this conservative cut in order to be sure that we can fully capture the variability of the AGNs, both through their bright and their faint phases. This choice is discussed further in Appendix A.   We also exclude objects in areas around the bad columns on the chips.  To accommodate small variations in the pointing of each field, these are chosen to be 100 pixels wide, leading to a loss of about $\sim 6\%$ of the field of view. \\
Our final sample consists of 28,096 AGNs with the total of 2.4 million photometric data-points. 
The PTF/iPTF survey also has a number of special fields which were configured for specific projects. Some of the selected AGNs therefore appear in different PTF-fields and CCD chips.
As our recalibration is chip and field specific (see Section \ref{sec:Cal}), we treat these duplicated AGNs as completely separate objects for calibration and analysis purposes. There are 3634 such objects, so our sample in fact consists of 31,730 calibrated light-curves. In Figures 1, 2 and 3 we present the distributions of the survey's observational cadence, the number of observations per light-curve and the total duration of the light-curves.  \\

At two points in our analysis, we also use light-curves from the SDSS Stripe 82 survey.  The first is when we estimate, and correct, the wavelength dependence of AGN variability. The second is to compare with the results of our PSD analysis. The Stripe 82 survey covered an area of 290 deg$^{2}$ and provides 5-band photometric measurements, with an average of 60 observations in each band per quasar. Photometric accuracy of the SDSS is excellent, with the photometric errors smaller than $\sigma$=0.018 mag in the $g$, $r$ and $i$ bands and 0.04 mag in $u$ and $z$ bands for sources with $r<$19 mag. We refer the reader to \cite{Mac12} for an extensive discussion of this survey and data products.   

\section{Re-Calibration of the PTF AGN Light Curves} \label{sec:Cal}

Although the initial calibration from  \cite{Ofe12} is available for all of the AGNs in our sample, we decided to recalibrate all of the AGN photometry in a way that was optimized to our scientific goals.  This led to a significant reduction in the systematic errors. A more detailed comparison of the calibrations is presented in Appendix B.  \\

In general, we use a relative calibration as we are primarily interested in the differences in the brightness of the quasars as a function of time rather than the absolute measurements. We broadly follow the procedure and philosophy in \cite{Lev11} and \cite{Ofe11}, for which we give a brief overview below. The main idea is to minimize the scatter in a set of reference objects, similar to the targeted quasars, that are available in the same images.  \\

We first identify, in each field, the five ``best" observations, i.e. epochs, simply in terms of the number of detected objects.  A set of potential ``reference" objects is then constructed to be all objects that (a) were observed in at least 4 out of these 5 best observations for a given field and chip combination, (b) have Sextractor \texttt{CLASS\_STAR} value larger than 0.9 and (c) do not have any Sextractor quality flags set in any of the observations \citep{Ber96}.  Potential reference objects that were selected in this fashion were cross checked against the AGN SDSS-DR7 catalog and known AGNs were removed.  We then assign an initial value of brightness to each of these reference objects, calculated as the mean instrumental magnitude of the object in these best observations. \\

For each object for which we want to create a calibrated light-curve, we then select a subsample of up to 50 reference objects which are in the same chip and same field.  These are chosen to be within 0.25 mag of the median magnitude of the five ``best'' observations of the target in question. If, as was usually the case, there are more than 50 objects that satisfy this criteria then we select the 50 spatially nearest ones (in the same chip) to the source that we are calibrating.  If there are fewer than ten available reference objects, the target is discarded. \\

Using these reference objects we then fit, using standard least-squares, the following relation for each of the targets:
\begin{equation}
m_{ij} \cong z_{i}+ \bar{m}_{j},
\end{equation}
where $m_{ij}$ is a $p \times q$ matrix that contains all the $p$ measured magnitudes of the $q$ reference objects, and $i$ is the epoch index and the $j$ is the index of reference object. From the results of the fit we can determine $\bar{m}_{j}$ which is the mean magnitude of the $j$th reference object and the $z_{i}$ which is the zero-point of the $i$th epoch observation \citep{Ofe12}. We apply the method iteratively, by removing reference objects which are highly variable ($\chi^{2}$/d.o.f.$\gt$ 3 for the constant magnitude fit) and adding ``cosmic error" term to all the measurement in the epoch for which the residuals are deemed to be too large (as suggested in \citealp{Ofe12}). \\

We then use the final zeropoints, $z_{i}$, determined for this target for each epoch, to calibrate the relative light-curve for the quasar. For each observation of a target at a given epoch, we also estimate the measurement errors $\sigma_{i}$
from the spread of the reference objects around their respective means.   We do not include specifically any color dependant terms in our calibration as we demonstrate that for stars there is no dependence of residuals on colour, as shown in Appendix B.  \\

Clearly, the estimate of the zero point $z_{i}$ at a given epoch will have uncertainties since it is based on data from a finite number of reference objects (maximum 50).  This estimate will introduce a spurious apparent variability to the target quasar that must be included in calculating the SF.  It is easy to show that the additional variance in the quasar variability will be equivalent to $\sigma^2 / (N-1)$ where N is the number of reference objects used and $\sigma^{2}$ is initially estimated variance. This additional variance can therefore be easily included in the final estimate of the measurement error.\\

Following the procedure in \cite{Sch10} we clean our resulting quasar light curves to remove spurious outliers. We do this by applying a 5-point median filter to the light-curves and removing individual observations which satisfy the criterion 
\begin{equation} \label{eq:clean}
|m_{i} - m_{m,i}|>0.3 \mbox{ mag} \times (\Delta t_{i,i-1}/100)^{1/2} + 5 \sigma_{i},
\end{equation}
where $m_{i}$ is the observed magnitude, $m_{m,i}$ is the resulting magnitude after applying 5-point median filter, $\Delta t_{i,i-1}$ is the temporal separation of $i$th and $i-1$ observation measured in days in the quasar rest frame and $\sigma_{i}$ is the estimated error of the $i$th observation. Given that the light-curves have large seasonal gaps, the time-dependent term ensures that we do not remove variability which is not spurious, but which might have occurred during the long observational break. We have compared our results with and without the time-dependent term in the Equation \eqref{eq:clean} and we find that our main conclusions are unchanged. This cleaning removes 0.1$\%$ of the data points. Visual inspection of the images associated with these discrepant data-points confirms that they were often compromised by satellite or aircraft trails \footnote{Re-calibrated data is available at \url{http://people.phys.ethz.ch/~caplarn/PTF/}}.   


\section{Analysis Methods} \label{sec:Met}

\subsection{Construction of the structure functions} \label{ConSF}

\begin{figure*}[htp]
    \centering
    \includegraphics[width=1.\textwidth]{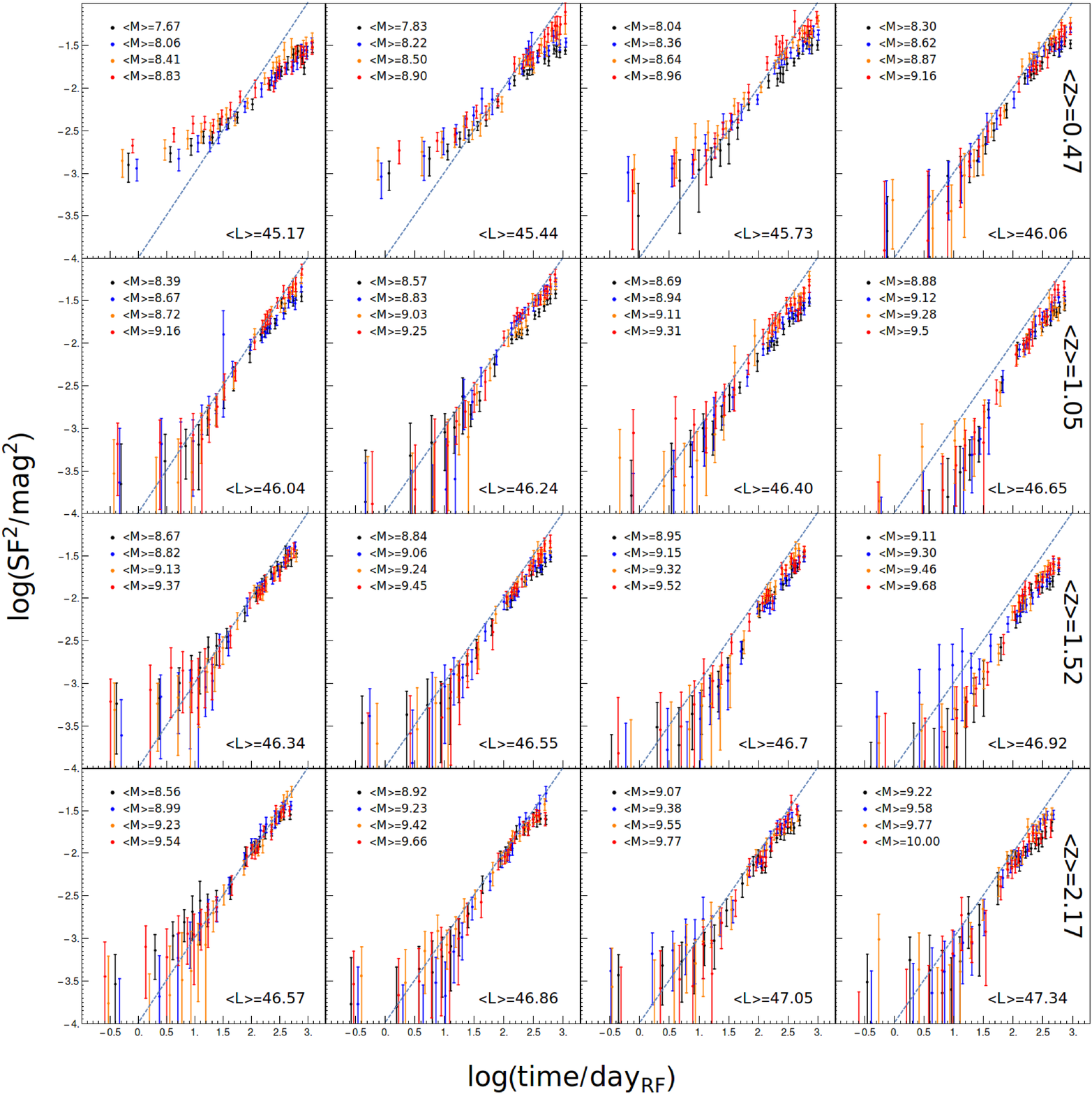}
    \caption{SF$^{2}$ split by redshift, luminosity and mass. Each row shows structure functions taken from the same redshift bin, as indicated on the right-hand side. Each panel within each row, then shows the structure functions of quasars in a given luminosity bin, which is then further split within each panel into black hole mass by denoting the points in  different colors. In each panel, we show the mean luminosity and the mean mass is given in the key, both in logarithmic units. The dashed line, which is the same on all of the panels, has a functional dependence of $-4+\log t$ and is given to help guide the eye in comparing panels.  Except at the lowest redshifts and luminosities, where the photometry could be affected by the host galaxy, the structure functions are consistent with zero as $\Delta t$ tends to zero.}
     \label{fig:AllzTogether}
\end{figure*}

In order to study the correlation of variability with the physical parameters of the quasars, we construct the ensemble structure functions.  We split our quasar sample into 64 bins defined by the physical properties of redshift, luminosity and black hole mass, as follows.  Firstly we split our sample into 4 redshift bins that each contain the same number of quasars. Each of these 4 redshift bins is then subdivided equally into 4 luminosity bins, and finally each of these bins is further divided into 4 mass bins. Mean values for redshift, mass and luminosity in each of these bins are shown in the Figure  \ref{fig:AllzTogether} and in the Table \ref{tab:bigtab}. Although the bins are different in size and are irregularly spaced in parameter space, this procedure ensures that each of the final 64 bins contains the same number of quasars and consequently also approximately the same number of photometric observations.   Since the uncertainty in the mean properties is driven by scatter within the population, this means that the uncertainty in the mean behaviour should be comparable in all of the bins.   \\

In each bin, we then construct the structure function using Equation \eqref{eq: SF}, modified as described in Section \ref{sec:Cal}. Firstly we transform all of the quasar observations into the quasar rest-frame, i.e. we divide all of the observation times by the time-dilation factor of (1+z), where $z$ is the redshift of the quasar. For each of the 64 bins, we then create the structure function in 20 time bins of time interval $\Delta t$, chosen so that each time bin contains the same number of pairs of observations. Outside of the $\Delta t$ associated with the seasonal gaps in the survey observations, the spacing of these bins is roughly equidistant in units of $\log \Delta t$, with a typical width of $\sim$ 0.13 dex. \\

This divides the whole survey, which consists of 250 million pairs of photometric observations, into 1280 bins of the structure function which contain, on average, 150,000 observational pairs each.  However, it should be realised that, within each structure function, the number of objects which are contributing to the different $\Delta t$ bins varies with $\Delta t$.  While time-bins describing short-term $\Delta t$ ($\sim 1$ day) contain typically 400-500 objects, because essentially all quasars in a given redshift-luminosity-mass bin will be present, the number contributing to the longest-term time bins drops to $\sim$ 150. This is a consequence of the inhomogeneous observing strategy in which objects are usually observed multiple times during the same or adjacent nights and only a subset of the sky was observed repeatedly during the full duration of the survey.         \\

Because our data sample is quite heterogeneous, with different quasars represented by a very different number of observations, there is some danger of the results being dominated by a small number of objects that were observed very often.  In the case of the SF, each light curve comprising $p$ data points will contribute $p^{2}$ times to the structure function.  Given that quasars display a wide range of variability properties, this could significantly bias our results, further exacerbating the problem that individual quasars may contribute to the SF$^{2}$ with different weights at different $\Delta t$.  To minimise this, when constructing the SF$^{2}$ in each of the time bins  $\Delta t$, we restrict the number of observational pairs from each quasar in a given  $\Delta t$ bin to be a maximum of 20 (randomly chosen) pairs. If the quasar does not contribute with 20 data-pairs to the particular time bin its contribution is discarded. This limits the weight given to the most densely sampled quasars.\\

To estimate the uncertainties in the SF$^{2}$  arising from the scatter in the population and from the variations in the observational data described in the previous paragraphs, we employ resampling procedures. We repeat the SF$^{2}$ construction procedure 100 times using $N$ randomly selected objects from the $N$ available objects, allowing repetition.  When an object is included twice, it is treated as ``new", e.g. with a different randomly chosen set of 20 observational pairs. We estimate the mean SF$^{2}$ and error on the SF$^{2}$ from the distribution of the results.\\

The quasar SF$^{2}$ constructed in this way show a significantly non-zero value of the SF$^{2}$ at short time scales, where the SF$^{2}$ should in principle converge towards zero for zero $\Delta t$.  For example, within a single night the SF$^2$ is typically $10^{-3.5}$, i.e. a residual variability of about 0.01-0.02 mag.  This could well be the signature of unrecognised systematic uncertainties, and similar samples of stars might be expected to show similar effects. We have therefore constructed control samples of stars which were matched in brightness to the AGN samples. These stellar samples were processed, using the above procedure, in exactly the same way as if they were quasars.   Of course, this control sample of stars could show genuine stellar variability effects.  This procedure is explained and discussed in detail in Appendix C, but we provide brief overview here. What is seen is that the stellar SF$^2$ is flat over a wide range of $\Delta t$, with a similar amplitude to that of the quasars at very small $\Delta t$.  The stellar SF$^2$ is independent of the colour of the stars.  It is therefore likely that the estimate of the observational errors on each photometric point has slightly underestimated the observational uncertainties in the survey (they appear to capture around 90\% of the variance). In order to account for this effect we simply subtract the stellar SF$^{2}$ from the initial AGN SF$^{2}$, to yield a final estimate of the AGN SF$^{2}$.  \\

We show the corrected SF$^{2}$ in Figure \ref{fig:AllzTogether}. We see that we are able to capture the variability for the vast majority of objects from shortest ($\lesssim$ 1 day) to very long (several years) scales . Except for a small excess variance at small $\Delta t$ at the lowest redshifts and luminosities, the SF$^{2}$ generally converge towards zero as $\Delta t$ approaches zero, as expected. The two lowest redshift and lowest luminosity bins may contain residual calibration errors, possibly due to host galaxy contributions. When conducting the analysis below we take special care to verify that none of our results is being driven by these two bins.  \\

\subsection{Wavelength correction}

When comparing the measurements of SF$^{2}$ of quasars of different properties, we would ideally compare them at the same rest-frame wavelength, given that variability is known to be wavelength dependent (e.g. \citealp{Mac10}). Since our data is only taken in a single (observed) band, this involves a redshift-dependent correction to the observed amplitudes, on top of correction to the timescales from the cosmological time-dilation.  This correction cannot be estimated from our data alone but can be estimated from the multi-wavelength quasar data from the SDSS Stripe 82 in \cite{Mac12} .\\

 We separate the available SDSS data into 5 redshift bins and then split each redshift bin into 5 luminosity bins. In each of the resulting 25 bins we estimate the ratios of the SF$^{2}$ in the $g-$ and $r-$ bands, and in the $r-$ and $i-$ bands. We estimate this ratio with all of the available data in the single bin as we assume that the color dependence of variability is independent of time. We do not use the $u$ and $z$ bands because of the larger errors associated with the photometric measurements in these bands.  This procedure, done in a single redshift bin, enables us to estimate the dependence of the SF$^2$ across the $gri$ wavelength range which spans the $r-$band data of the (i)PTF.   Splitting the redshift bin into different luminosity bins allows us to check if this wavelength dependence changes with luminosity, but we do not in fact detect any such luminosity trends and can treat these five luminosity bins as independent estimates of a luminosity-independent relation. \\
 
\begin{figure}[htp]
    \centering
    \includegraphics[width=.49\textwidth]{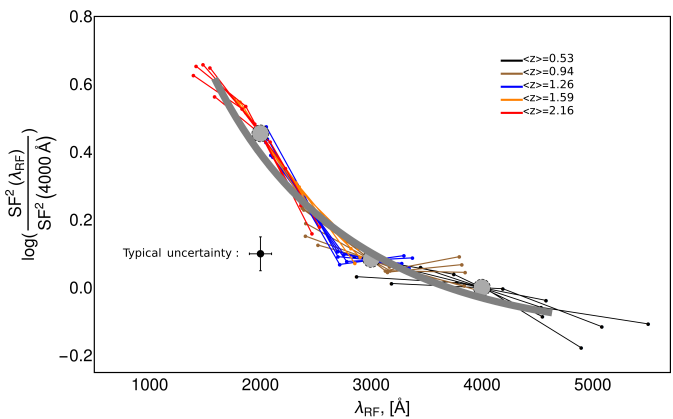}
    \caption{The redshift ladder constructed to determine the dependence of variability as a function of wavelength. We estimate the ratio of excess variance (SF$^{2}$) between the g-r band and r-i band in the SDSS Stripe 82 in different redshift bins and then demand that these estimates agree at the fixed points of 2000, 3000, and 4000 \AA. 
The grey line, which we show in the wavelength range that is probed in this survey, shows a fit to the data (Equation  \ref{eq:KCor}) which we use in further analysis to characterize the dependency of variability with wavelength.   See text for more details. }
     \label{fig:KCorrection}
\end{figure} 
 
In order to connect these single-redshift estimates across a range of redshift, we construct a redshift ladder, as shown in the Figure \ref{fig:KCorrection}. We first force the $gri$ SF$^2$ curves from the five luminosity bins at the lowest redshift to be coincident at a rest-frame 4000 $\text{\AA}$ by making small vertical adjustments.  We then require the SF$^2$ variance curves from the next two redshift bins to overlap with each other and with the ones from the lowest redshift bin at the shorter rest-frame wavelength of 3000 $\text{\AA}$.  Finally, we can add the two highest redshift bins, now normalising at 2000 $\text{\AA}$.   We then fit a second order polynomial to these normalized estimates of SF$^2$ at different wavelengths to infer:
\begin{equation}\begin{split} \label{eq:KCor}
\log \frac{\mbox{SF}^{2}(\lambda_{rf})}{\mbox{SF}^{2}(4000 \text{\AA})}&=-0.03 - 0.815  \log \left(\frac{\lambda}{4000 \text{\AA}}\right)  \\
&+ 1.989 \log\left( \frac{\lambda}{4000 \text{\AA}}\right) ^{2}.
\end{split}\end{equation}
We will use the function in Equation \eqref{eq:KCor} to renormalise all of our estimates of variability at different redshifts to the value expected at 4000 $\text{\AA}$. 
This choice of rest-frame wavelength corresponds to what is observed within (i)PTF for $z\sim1$ quasars - typical of our sample.
We note that our prescription is consistent with other estimates of the wavelength dependence of quasar variability from the same dataset (\citealp{Mac10},  \citealp{Meu11}, \citealp{Zuo12}).\\

\subsection{Fitting the structure function}

As was seen in Figure \ref{fig:AllzTogether}, the SF$^{2}$ rise uniformly with time, with slopes similar to that expected from a random walk and with possible breaks at long time-scales in some cases. We therefore model the ensemble SF$^{2}$, constructed in each bin of quasar properties and after applying the wavelength correction derived above, with the functional form of a broken power law. Namely we use 
\begin{equation} \label{eq:SFeq}
\mbox{SF}^{2} (\Delta t) = \left\{
        \begin{array}{ll}
          \phi_{0}+ \phi_{1} \cdot (\Delta t)^{g_{1}} & \quad \Delta t \leq \Delta t_{br} \\
    \left(  \phi_{0}+  \phi_{1} \cdot (\Delta t_{br})^{g_{1}}\right) \cdot  \left(  \frac{\Delta t}{\Delta t_{br}} \right)^{g_{2}}  & \quad \Delta t >\Delta t_{br} 
        \end{array}
    \right.
\end{equation}
where the  $\Delta t_{br}$ is the timescale associated with a break in the structure function. We fit the SF$^2$ function with a double power law to allow for potential changes of slope at long timescales which can be caused by either a real flattening or by statistical fluctuations in the sample (remembering that not all quasars contribute at the longest timescales). The term $\phi_{0}$ in Equation \eqref{eq:SFeq} allows for the possibility of a time-independent term associated with lingering systematic effects in the calibration. All of the inferred parameters are shown in Table \ref{tab:bigtab}.\\

\begin{deluxetable*}{ccclllllr}  
\tablecolumns{15}
\tablewidth{0pt} 
\tablecaption{Fit with Equation \eqref{eq:SFeq}, table of parameters  }
\tablehead{   
\colhead{$\left\langle z  \right\rangle$} &
\colhead{$\left\langle \log L  \right\rangle$} &
\colhead{$\left\langle \log M  \right\rangle$} &
  \colhead{$\phi_{0}$} &
  \colhead{$\phi_{1}$} &
  \colhead{$g_{1}$} &
  \colhead{$g_{2}$} &
  \colhead{$t_{br}$}&
  \colhead{$\chi^{2}/\mbox{d.o.f}$}  
}
\startdata
0.26&45.15&7.67&-2.88$\pm$0.12&-3.94$\pm$0.24&0.83$\pm$0.098&0.242$\pm$0.11&2.70$\pm$0.088&0.75
\\
0.28&45.17&8.06&-3.02$\pm$0.15&-4.08$\pm$0.26&0.93$\pm$0.12&0.531$\pm$0.071&2.18$\pm$0.085&0.37
\\
0.29&45.17&8.41&-2.77$\pm$0.10&-4.29$\pm$0.31&1.04$\pm$0.13&0.348$\pm$0.14&2.49$\pm$0.14&0.42
\\
0.30&45.18&8.83&-2.67$\pm$0.083&-3.80$\pm$0.25&0.78$\pm$0.10&0.461$\pm$0.081&2.25$\pm$0.12&0.66
\\
0.38&45.42&7.83&-3.68$\pm$0.61&-3.83$\pm$0.26&0.86$\pm$0.11&0.393$\pm$0.074&2.35$\pm$0.083&0.85
\\
0.40&45.44&8.22&-4.37$\pm$0.71&-3.34$\pm$0.13&0.67$\pm$0.057&0.327$\pm$0.10&2.54$\pm$0.15&0.57
\\
0.40&45.44&8.50&-2.77$\pm$0.094&-4.78$\pm$0.37&1.29$\pm$0.17&0.699$\pm$0.11&2.18$\pm$0.098&0.81
\\
0.40&45.44&8.90&-2.82$\pm$0.14&-4.01$\pm$0.20&0.96$\pm$0.082&0.455$\pm$0.20&2.73$\pm$0.19&0.56
\\
0.51&45.71&8.04&-5.66$\pm$1.8&-4.04$\pm$0.20&0.96$\pm$0.11&0.519$\pm$0.081&2.33$\pm$0.19&0.70
\\
0.54&45.72&8.36&-10.7$\pm$1.4&-3.68$\pm$0.11&0.82$\pm$0.052&0.391$\pm$0.11&2.53$\pm$0.098&1.1
\\
0.56&45.74&8.64&-15.8$\pm$2.2&-3.58$\pm$0.13&0.80$\pm$0.058&0.387$\pm$0.14&2.60$\pm$0.11&1.1
\\
0.55&45.74&8.96&-3.68$\pm$0.58&-4.15$\pm$0.29&1.12$\pm$0.15&0.453$\pm$0.14&2.36$\pm$0.16&0.46
\\
0.64&46.01&8.30&-4.33$\pm$0.78&-4.27$\pm$0.21&1.04$\pm$0.088&0.266$\pm$0.15&2.60$\pm$0.082&0.34
\\
0.67&46.05&8.62&-4.72$\pm$1.6&-4.36$\pm$0.26&1.13$\pm$0.12&0.482$\pm$0.13&2.43$\pm$0.11&0.73
\\
0.69&46.08&8.87&-7.53$\pm$0.85&-4.86$\pm$0.25&1.43$\pm$0.13&0.695$\pm$0.082&2.15$\pm$0.078&0.44
\\
0.66&46.11&9.21&-4.30$\pm$0.50&-4.45$\pm$0.21&1.19$\pm$0.11&0.751$\pm$0.074&2.10$\pm$0.080&0.28
\\
0.90&45.99&8.39&-4.43$\pm$0.60&-4.20$\pm$0.18&1.03$\pm$0.079&0.364$\pm$0.19&2.57$\pm$0.12&0.42
\\
0.91&46.02&8.67&-4.19$\pm$0.57&-4.05$\pm$0.17&0.98$\pm$0.086&0.676$\pm$0.081&2.25$\pm$0.12&0.56
\\
0.95&46.05&8.87&-4.29$\pm$0.56&-4.60$\pm$0.30&1.29$\pm$0.15&0.719$\pm$0.090&2.16$\pm$0.11&0.29
\\
0.97&46.07&9.16&-11.3$\pm$1.3&-4.37$\pm$0.23&1.16$\pm$0.11&0.635$\pm$0.17&2.45$\pm$0.17&0.43
\\
1.02&46.24&8.58&-3.69$\pm$0.44&-4.84$\pm$0.28&1.34$\pm$0.14&0.787$\pm$0.053&2.08$\pm$0.049&0.37
\\
1.03&46.25&8.84&-4.51$\pm$0.67&-4.95$\pm$0.28&1.49$\pm$0.14&0.707$\pm$0.057&2.07$\pm$0.050&0.35
\\
1.06&46.25&9.03&-8.03$\pm$1.0&-4.35$\pm$0.22&1.11$\pm$0.10&0.530$\pm$0.20&2.56$\pm$0.15&0.35
\\
1.05&46.25&9.25&-5.89$\pm$0.65&-4.63$\pm$0.28&1.32$\pm$0.14&0.692$\pm$0.095&2.19$\pm$0.096&0.30
\\
1.08&46.39&8.69&-3.93$\pm$0.66&-4.56$\pm$0.37&1.16$\pm$0.18&0.746$\pm$0.087&2.15$\pm$0.16&0.46
\\
1.10&46.40&8.94&-5.40$\pm$1.1&-4.31$\pm$0.19&1.05$\pm$0.086&0.455$\pm$0.17&2.49$\pm$0.092&0.26
\\
1.12&46.41&9.11&-7.75$\pm$0.84&-4.89$\pm$0.32&1.40$\pm$0.16&0.794$\pm$0.12&2.15$\pm$0.12&0.57
\\
1.11&46.41&9.31&-6.46$\pm$0.47&-4.08$\pm$0.24&1.02$\pm$0.12&0.481$\pm$0.14&2.30$\pm$0.12&0.81
\\
1.09&46.59&8.88&-14.0$\pm$0.92&-5.07$\pm$0.27&1.41$\pm$0.15&0.631$\pm$0.080&2.19$\pm$0.10&0.26
\\
1.14&46.61&9.12&-6.11$\pm$1.3&-5.03$\pm$0.19&1.35$\pm$0.089&0.664$\pm$0.17&2.46$\pm$0.098&0.30
\\
1.15&46.65&9.28&-9.72$\pm$1.0&-4.68$\pm$0.31&1.20$\pm$0.14&0.499$\pm$0.17&2.42$\pm$0.085&0.58
\\
1.16&46.69&9.51&-10.9$\pm$1.7&-4.84$\pm$0.19&1.28$\pm$0.091&0.889$\pm$0.20&2.43$\pm$0.28&0.38
\\
1.45&46.29&8.67&-7.20$\pm$2.2&-3.83$\pm$0.21&0.90$\pm$0.096&0.419$\pm$0.13&2.39$\pm$0.12&0.35
\\
1.45&46.34&8.92&-3.64$\pm$0.36&-4.72$\pm$0.31&1.32$\pm$0.15&0.770$\pm$0.080&2.09$\pm$0.068&0.34
\\
1.44&46.37&9.13&-4.11$\pm$0.81&-4.08$\pm$0.22&0.98$\pm$0.10&0.365$\pm$0.17&2.60$\pm$0.15&0.59
\\
1.45&46.37&9.37&-4.21$\pm$0.87&-4.26$\pm$0.23&1.08$\pm$0.10&0.654$\pm$0.14&2.24$\pm$0.17&0.64
\\
1.51&46.55&8.84&-16.7$\pm$1.7&-4.27$\pm$0.19&1.07$\pm$0.094&0.602$\pm$0.12&2.22$\pm$0.11&0.21
\\
1.49&46.55&9.06&-10.4$\pm$0.82&-4.23$\pm$0.18&1.04$\pm$0.082&0.467$\pm$0.20&2.47$\pm$0.085&0.46
\\
1.47&46.56&9.24&-10.2$\pm$0.92&-4.28$\pm$0.16&1.08$\pm$0.071&0.451$\pm$0.19&2.57$\pm$0.064&0.38
\\
1.50&46.55&9.45&-8.60$\pm$1.4&-4.79$\pm$0.28&1.34$\pm$0.14&0.880$\pm$0.089&2.11$\pm$0.071&0.20
\\
1.55&46.70&8.95&-10.1$\pm$0.82&-4.65$\pm$0.21&1.22$\pm$0.11&0.759$\pm$0.096&2.16$\pm$0.095&0.24
\\
1.53&46.70&9.15&-5.22$\pm$0.72&-4.49$\pm$0.16&1.10$\pm$0.069&0.480$\pm$0.22&2.68$\pm$0.15&0.79
\\
1.55&46.71&9.32&-8.35$\pm$1.0&-5.06$\pm$0.31&1.45$\pm$0.15&0.850$\pm$0.086&2.13$\pm$0.11&0.30
\\
1.55&46.71&9.52&-12.2$\pm$0.85&-4.07$\pm$0.15&0.96$\pm$0.065&0.408$\pm$0.19&2.64$\pm$0.13&0.71
\\
1.59&46.89&9.11&-5.63$\pm$1.1&-5.09$\pm$0.38&1.38$\pm$0.18&0.492$\pm$0.17&2.25$\pm$0.084&0.26
\\
1.59&46.89&9.30&-4.27$\pm$1.5&-4.43$\pm$0.39&1.06$\pm$0.18&0.591$\pm$0.19&2.41$\pm$0.20&0.61
\\
1.60&46.94&9.46&-4.36$\pm$0.51&-4.69$\pm$0.24&1.19$\pm$0.11&0.506$\pm$0.20&2.48$\pm$0.087&0.70
\\
1.59&46.99&9.68&-6.20$\pm$1.0&-5.15$\pm$0.23&1.46$\pm$0.12&0.649$\pm$0.11&2.18$\pm$0.074&0.28
\\
2.04&46.48&8.56&-8.71$\pm$1.8&-3.59$\pm$0.14&0.80$\pm$0.066&0.324$\pm$0.17&2.49$\pm$0.16&0.33
\\
1.96&46.57&8.99&-4.31$\pm$0.53&-4.00$\pm$0.17&1.01$\pm$0.080&0.426$\pm$0.17&2.44$\pm$0.11&0.37
\\
1.97&46.61&9.23&-6.80$\pm$1.1&-4.30$\pm$0.20&1.12$\pm$0.090&0.616$\pm$0.24&2.57$\pm$0.21&0.42
\\
2.03&46.64&9.54&-3.86$\pm$0.61&-4.10$\pm$0.24&1.02$\pm$0.11&0.343$\pm$0.17&2.47$\pm$0.086&0.40
\\
2.05&46.85&8.92&-6.65$\pm$0.84&-4.39$\pm$0.22&1.18$\pm$0.11&0.420$\pm$0.13&2.20$\pm$0.079&0.33
\\
1.98&46.85&9.23&-4.86$\pm$0.57&-4.69$\pm$0.23&1.29$\pm$0.12&0.732$\pm$0.22&2.37$\pm$0.17&0.53
\\
2.06&46.86&9.42&-10.1$\pm$1.3&-4.00$\pm$0.17&0.98$\pm$0.085&0.522$\pm$0.19&2.40$\pm$0.18&0.44
\\
2.05&46.88&9.66&-6.57$\pm$0.73&-4.40$\pm$0.26&1.17$\pm$0.13&0.484$\pm$0.15&2.25$\pm$0.097&0.31
\\
2.22&47.04&9.07&-5.57$\pm$1.6&-4.29$\pm$0.26&1.04$\pm$0.12&0.423$\pm$0.19&2.45$\pm$0.13&0.95
\\
2.17&47.05&9.38&-3.29$\pm$0.16&-4.84$\pm$0.30&1.35$\pm$0.14&0.633$\pm$0.15&2.28$\pm$0.065&0.34
\\
2.13&47.05&9.57&-4.41$\pm$0.99&-4.61$\pm$0.39&1.27$\pm$0.19&0.522$\pm$0.13&2.19$\pm$0.089&0.33
\\
2.21&47.08&9.77&-6.41$\pm$1.5&-4.48$\pm$0.21&1.17$\pm$0.095&0.549$\pm$0.21&2.43$\pm$0.11&0.66
\\
2.56&47.28&9.22&-8.46$\pm$0.73&-4.57$\pm$0.31&1.22$\pm$0.16&0.548$\pm$0.12&2.05$\pm$0.049&0.44
\\
2.30&47.27&9.58&-4.93$\pm$0.85&-4.53$\pm$0.21&1.18$\pm$0.092&0.582$\pm$0.23&2.41$\pm$0.12&0.42
\\
2.44&47.31&9.77&-12.9$\pm$1.1&-4.61$\pm$0.28&1.25$\pm$0.14&0.707$\pm$0.16&2.20$\pm$0.14&0.78
\\
2.61&47.40&10.0&-7.85$\pm$1.3&-4.67$\pm$0.30&1.25$\pm$0.16&0.484$\pm$0.18&2.19$\pm$0.10&0.32
\\
\enddata
\label{tab:bigtab}
\end{deluxetable*}

We note that while we use a broken power law to describe our SF$^2$ functions by ensuring a good fit, we do not think it is safe to ascribe physical meaning or significance to the timescales of the possible breaks which we always find on the scales comparable to the total length of the survey. The main reason for this is that, as noted above, different quasars will contribute with different weights to the SF$^2$ at different $\Delta t$, because of the  heterogeneous temporal coverage of the (i)PTF survey in different fields.  Only a relatively small number of quasars have photometric data covering the full length of the survey.  In quite general terms, it is clear that features in the SF on timescales comparable to the length of the survey should be treated with some caution because of limited sampling at these timescales, even for individual objects (e.g. \citealp{Emm10}).

\subsection{Construction and testing of PSDs} \label{sec:DecTestPSD}

Power spectrum analysis presents a powerful alternative to structure function analysis. In this section we describe the analysis of the data using this independent method. While the SF$^{2}$ analysis was used on the ensemble of objects, the PSD analysis can be only used on individual light-curves. We will be using the  code presented in \cite{Kel14} that estimates the PSD of a light curve by modelling it as a continuous auto-regressive moving average (CARMA) process. This method is suitable for use even when using data with non-uniform time sampling with large gaps, as in (i)PTF. The tool includes an adaptive MCMC sampler, maximum likelihood estimators and the basic tools for representing the final results of the estimate. We refer the reader to \cite{Kel14} for a detailed explanation of the procedure and the code. Previous examples of the AGN variability studied using this algorithm can be found in \cite{Ede14} and \cite{Sim16}. \\

\begin{figure}[htp] 
    \centering
    \includegraphics[width=.49\textwidth]{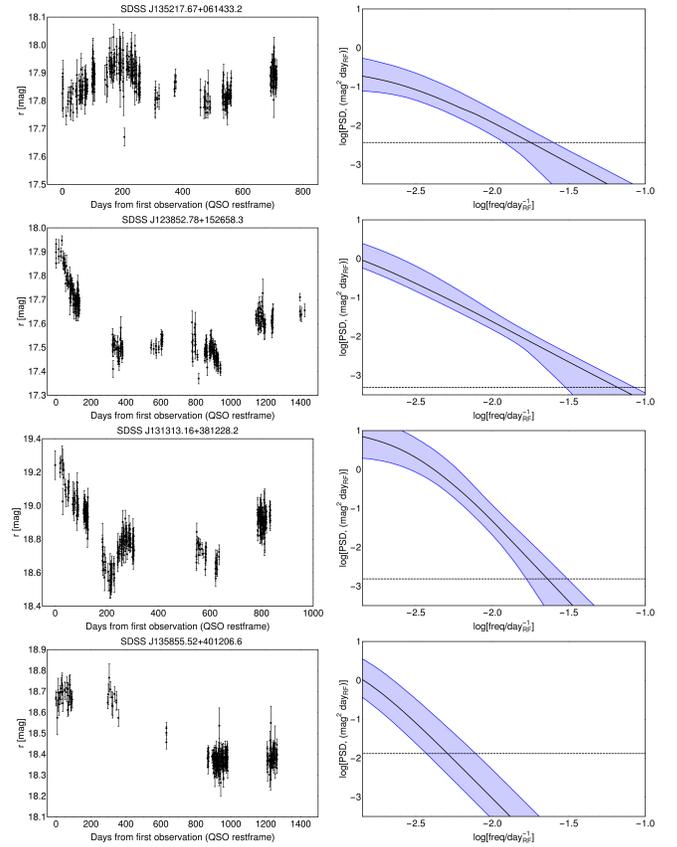}
    \caption{Four representative light curves (left) and the resulting PSD spectra (right) obtained using the code presented in \cite{Kel14}. In the latter, the solid lines show the maximum-likelihood estimate, while the shaded regions shows the 95\% confidence interval. The dashed lines show the approximate level of noise in the photometric measurements. All of the right hand panels have the same scaling to ease comparison. The upper two panels show examples of quasars that are well modelled with with power-law PSD with slope of $\alpha \sim 2$, while the lower two panels shows two quasars with a steeper PSD, $\alpha \sim 3$. 
    }
    \label{fig:LCToPSD}
\end{figure} 

For setting up the algorithm we use the standard procedure described in \cite{Kel14} and further elaborated in \cite{Sim16}. We let the algorithm freely choose the order of the CARMA(P,Q) process on the grid of values P=1,2,3 and Q=0,1,2.  It was not found to be necessary to include higher order processes as the light curves were adequately modelled in the vast majority of cases. After choosing the optimal CARMA(P,Q) process we then run the sampler for 75,000 iterations. We show examples of the results in Figure \ref{fig:LCToPSD}. \\ 

 Following testing of this algorithm, we restricted the analysis to well-sampled, extended light-curves. Specifically, we include only light curves with more than 60 data points and more than 300 days separation between the first and last observations. Light-curves with lower quality tend to perform poorly and often produce results which are of little value due to the large inferred errors on the estimates. Imposing these conditions reduces our sample ten-fold to $\sim$ 2,200 quasars.  \\

We performed extensive simulations to see how well we are able to infer the parameters of the light curves with this particular method in this particular data-set. In order to do so we simulate light curves with the same cadence and observational errors as in the real data-set. We use the algorithm from \cite{Tim95} to generate light curves with the PSDs which are described with single power-law slope of $\alpha=1.75,2,2.25,2.5,2.75,3$.  We then try to recover the slope by fitting a power law to the PSD estimated with the CARMA procedure. In order to eliminate red-noise leak effects (see e.g. \citealp{Emm10}) we generate ten times longer light curves than needed and then select a random segment having the desired length. To match the observed ensemble behaviour of our sample (see Section \ref{sec:SF}) we normalize the variability by increasing or decreasing the simulated variability so that the SF$^{2}$ for the whole sample of simulated AGNs at 100 days is equal to 0.01 mag$^{2}$. \\

When fitting the PSD we only fit the values which are above the estimated level of measurement noise in the data. Following \cite{Kel14} and \cite{Sim16}, the measurement noise is set to 2 \rm{med}$ (\delta t) \cdot \rm{med}( \sigma^{2})$ where med $(\delta t )$ is the median cadence and med$( \sigma^{2})$  is the median of the measured noise variance. We tested also the results when using the means of the cadence and variance instead of the medians, but find that choosing medians reproduces better the input values in the simulation.   \\

\begin{figure}[htp]
    \centering
    \includegraphics[width=.49\textwidth]{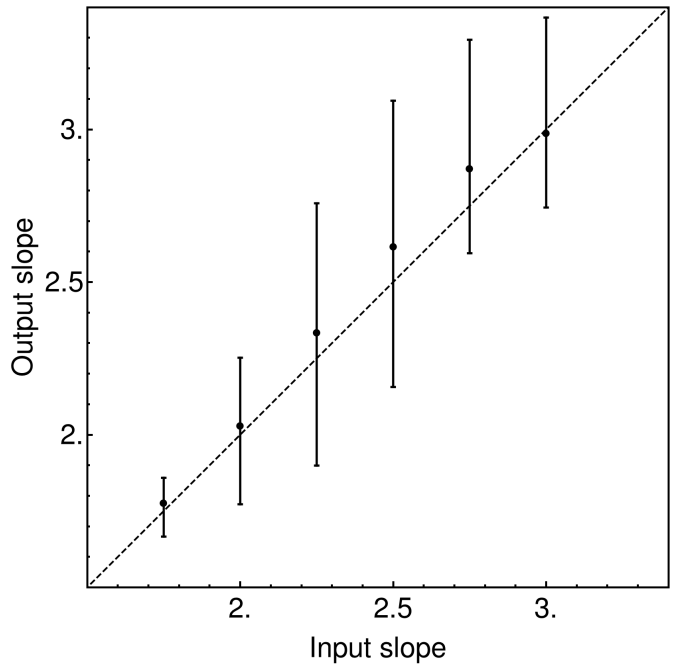}
    \caption{Result of the simulations in which we attempt to recover the slope of the PSD of a set of quasars whose mock photometry was generated using a power-law PSD of known slope, as described in the text.   The mock observations are constructed to have the same errors and cadence etc as the actual data.  While the PSD analysis recovers the mean slope well, there is significant dispersion in the output for a given input slope.  The error bars represent the 25 and 75 percentile of the distribution of the resulting output slopes.}
     \label{fig:InputOutput}
\end{figure}

\begin{figure}[htp] 
    \centering
    \includegraphics[width=.49\textwidth]{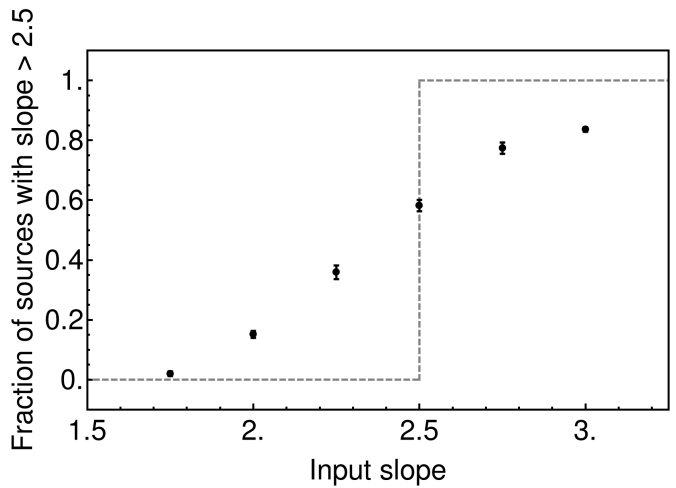}
    \caption{An alternative way to show the results of the simulations in Figure \ref{fig:InputOutput} is to plot the fraction of sources that are inferred to have steep slope $\alpha > $2.5, where PSD $\propto f^{-\alpha}$, including sampling errors derived from bootstrapping the mock sample (as with the real data). Ideally, we would recover the dashed line step function, which represents a case in which the input and the output slopes would agree perfectly.  Uncertainties in determining the slope produce a blurring of this.}
    \label{fig:InputOutput2}
\end{figure}

We show the results of these simulations in Figure \ref{fig:InputOutput}. The error bars represent the 25 and 75 percentiles of the distribution of the recovered parameters from the population and not the inferred errors for a single object. Although the spread is large the basic trend is recovered. The large spread highlights the fact that the estimated parameters for a single light-curve are quite uncertain and that conclusions can be only made by using a statistically significant number of objects. We also note that the slopes are better recovered for slopes around two (corresponding to random walk), than for steeper slopes.  This reflects the fact that for intermediate values of the slope ($\sim$ 2.5), the algorithm tends to sometimes run away and converge to extremal values of either 2 or 4 (see also Figure B.1. from \citealp{Sim16}). In Figure \ref{fig:InputOutput2}, we show an alternative way of representing the results, which partially alleviates these effects. Here we represent the fraction of sources that are inferred to have a power law slope steeper than $\alpha >$  2.5. In the ideal case we would want to see a clear step function, with a step at the input value of $\alpha =2.5$ (as represented with a dashed line in the Figure).  In practice, we do recover the basic trends, but do not reproduce the step function exactly. The error bars in the Figure are derived by bootstraping the sample and as such represent the statistical uncertainty in the number of quasars with PSD slopes larger than 2.5, given the number of quasars in the sample, which is set to be representative of the actual data.\\

We remark that we tried fitting the PSDs from both the data and the simulated light-curves with a broken power-law fit, but found the results to be unreliable. Specifically, when running our simulations, which as input had pure power-law PSDs , we found that the inferred (rest-frame) time scale of the break was perfectly correlated with the redshift i.e. effectively with the (rest-frame) duration of the observations. This is due to the fact that the algorithm is modelling the PSD as a weighted sum of Lorentzian functions \citep{Kel14} which means that the function which is used to model the PSD must have a turnover at some point. The model is forced to locate this turnover at or near to the minimal frequency available in the observations.  We see exactly the same effect when fitting the real data, with a clear dependence on redshift but not on mass or luminosity, from which we conclude that this is an artificial effect.   In general, we find that with the double power-law, the algorithm tends to overestimate the steepness of the low-frequency slopes and to give relatively unreliable estimates of the high-frequency slope. Therefore we conclude that the details of the PSD are unreliable, and that we are only able to estimate, at least with the (i)PTF data, the global shape of the PSD for statistical samples of objects.

\section{Results} \label{sec:SF}

\subsection{The amplitude and the timescale of variability from the SF analysis} \label{sec:Timescale}

 \begin{figure*}[htp]
    \centering
    \includegraphics[width=.99\textwidth]{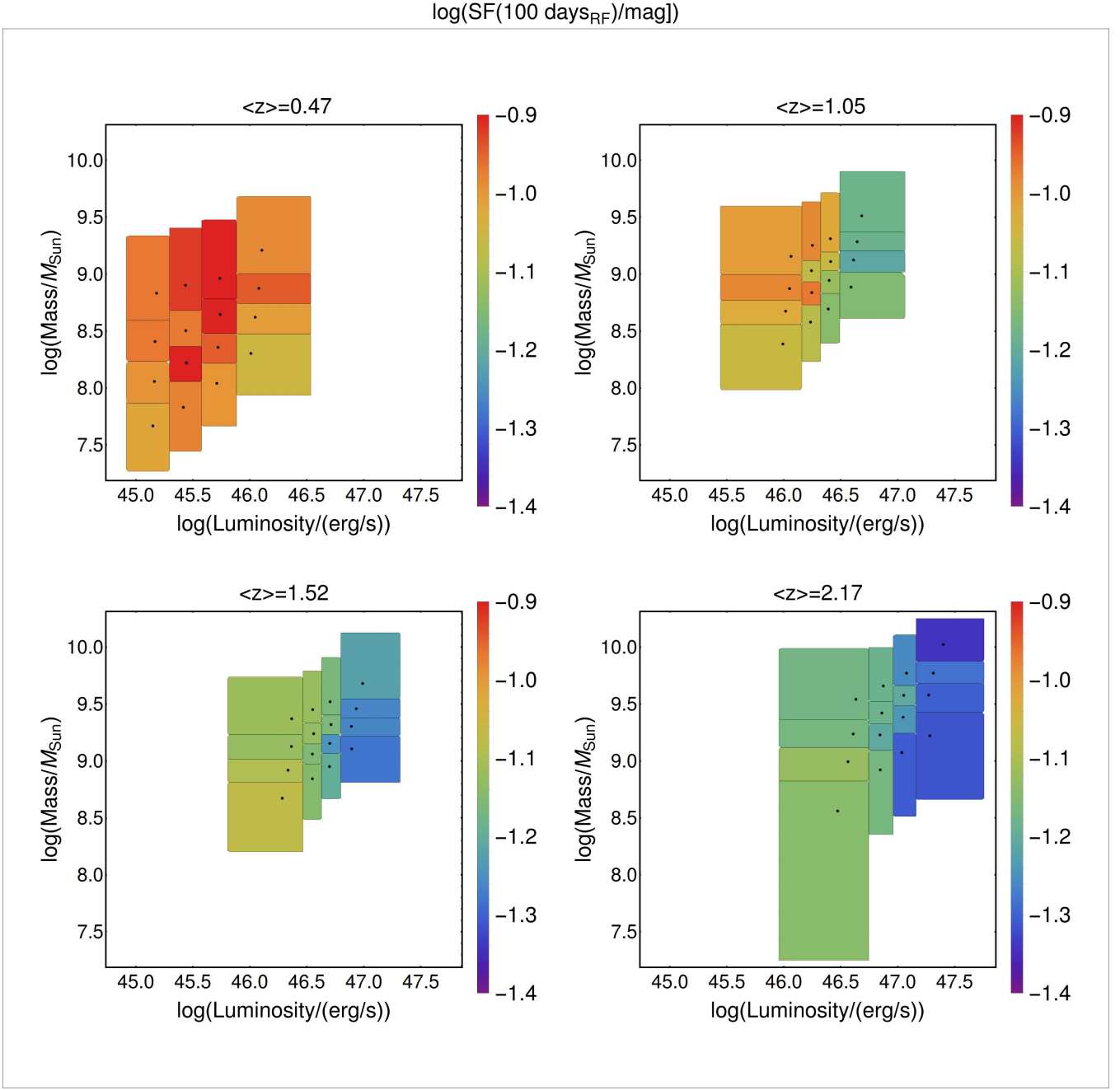}
    \caption{Value of the $\log$(SF/mag) at a time scale of 100 days, in the quasar restframe as a functions of the black hole mass and the luminosity, as observed in the four redshift intervals (the four panels). The dots inside each bin show the mean values of mass and luminosity for the quasars in that bin. A clear anti-correlation with luminosity can be observed, with more luminous objects varying less. This dependence is seen in individual redshift bins and also across the whole redshift range.   }
     \label{fig:SF100}
\end{figure*}

\begin{figure}[htp]
    \centering
    \includegraphics[width=.49\textwidth]{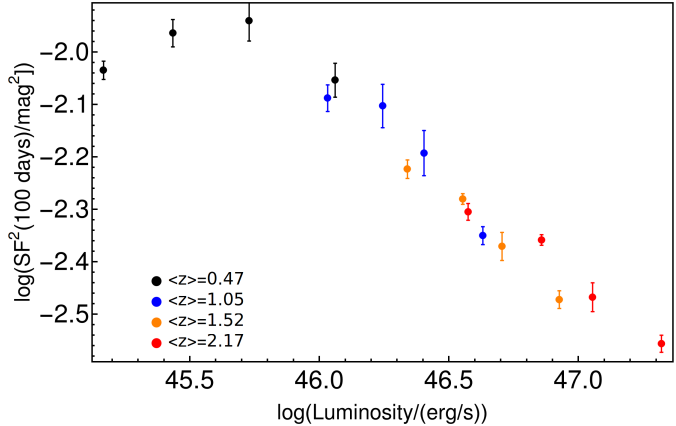}
    \caption{The dependence of the SF$^{2}$, measured at 100 days, as a function of the luminosity $L$. The strong anti-correlation with luminosity, and the weak or absent dependence on redshift, are obvious. }
     \label{fig:LPlot}
\end{figure}

 \begin{figure*}[htp]
    \centering
    \includegraphics[width=.99\textwidth]{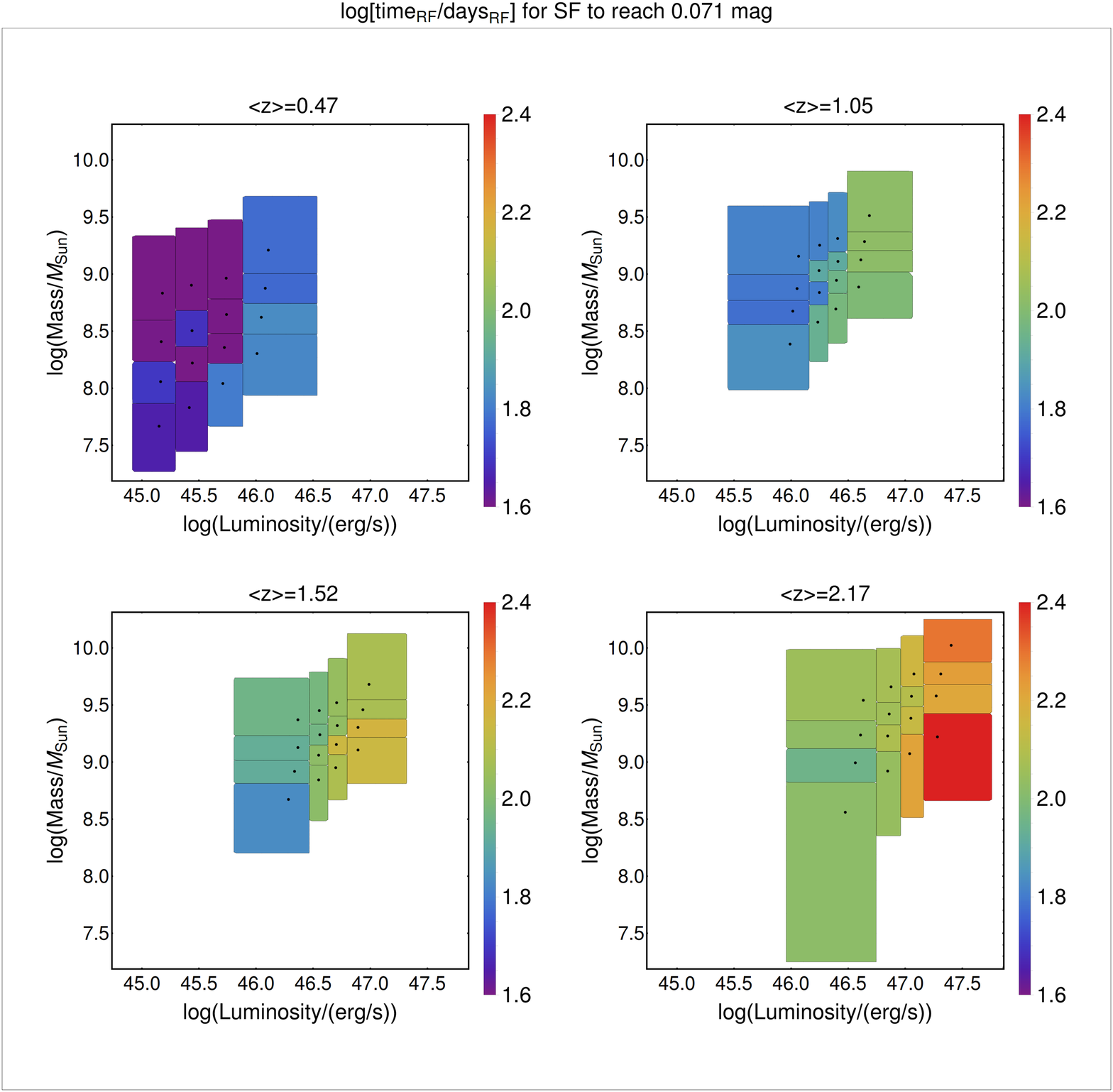}
    \caption{The logarithm of the rest-frame time-scale, in days, at which the ensemble structure function reaches 0.071 mag. The dots inside the bins show the mean value of mass and luminosity for the objects in the bin. This figure contains similar information to that in the Figure   \ref{fig:SF100} but offers a different interpretation of the variability properties. The clear dependence with luminosity can be observed, with more luminous objects needing more time to achieve the same level of variability. This dependence is seen in individual redshift bins and also across the whole redshift range.}
     \label{fig:TimeTo007}
\end{figure*} 

In Figure \ref{fig:SF100} we show the amplitude of variability at a time-scale of 100 days (in the quasar rest frame) for the 64 bins in redshift, luminosity and black hole mass. This timescale is chosen as the associated level of variability is well above the noise level of the survey (see Figure \ref{fig:AllzTogether}) and because it lies well within the most densely sampled parts of the survey (significantly shorter than the total length of the survey).  \\

A strong anti-correlation is seen between the amplitude of variability and luminosity - more luminous sources are less variable. This effect is visible in every redshift bin and is even more obvious when comparing different redshift panels, as the distribution of the sample shifts in luminosity. We explicitly show this result in Figure \ref{fig:LPlot}, where we have suppressed the mass information and show the variability amplitude at 100 days (rest-frame) in bins defined in redshift and luminosity. We observe the clear and tight anti-correlation of variability with luminosity, i.e. the effect that we already discussed above. We note that the flattening on the low-luminosity side is possibly spurious - as mentioned before, the largest residuals in our estimation of SF$^{2}$ are in the lowest luminosity and lowest redshift bins. Apart from that, the trend with luminosity is well established and continuous across redshift bins. This clear dependence with luminosity has been seen before several times in smaller surveys (e.g. \citealp{Van04}, \citealp{Bau09}, \citealp{Mac10}) but is very clearly seen in the very large data set presented here.  A comparison of the panels in Figure \ref{fig:SF100} or examination of Figure \ref{fig:LPlot}  shows that the variability at a given luminosity is more or less constant with redshift.  Although the dependence on luminosity is obvious, any dependence with black hole mass is much less pronounced. \\ 

In Figures \ref{fig:SF100} and \ref{fig:LPlot} we followed the conventional practice of considering the amplitude of variability at a given (rest-frame) timescale.  A more or less equivalent diagram results from what we might suspect is a possibly more physically intuitive representation, which is to consider the timescale $\tau$ at which the variability reaches some given amplitude. This is shown in Figure \ref{fig:TimeTo007}.  We set this amplitude to be 0.071 mag, equivalent to the SF$^2$ reaching 0.005 mag$^{2}$. This value is similarly  chosen as it is far above the noise level of the survey and because this level of variability is reached between 50 to 250 days, ensuring that our estimates are again within very densely sampled parts of the survey.  It is to be also noted that this timescale is of the order of the typical dynamical time-scales expected for the AGNs in our sample (e.g. \citealp{Net13}). \\

Formally, for each of our AGN samples we search for $\tau$  that is the solution to the simple equation
\begin{equation}
\mbox{SF}^{2} (\tau_{C}) = C \mbox{ mag$^{2}$}.
\end{equation}
where $C$=0.005. We note that for short enough times, where SF$^{2}$ is fully described with a single power law (see Equation \ref{eq:SFeq}) this connection can be expressed as 

\begin{equation}
\tau_{C} = \left( \frac{C}{\phi_{1}}\right)^{g_{1}}. 
\end{equation}

Not surprisingly, the two Figures \ref{fig:SF100} and \ref{fig:TimeTo007} are similar.  Again, we see a clear positive correlation between luminosity and $\tau_{C}$, and little variation with black hole mass or redshift (recall that the analysis is performed in the rest-frame).  The statement that more luminous sources vary on longer timescales is largely equivalent to the previous statement that they are less variable at a given timescale.  \\

To investigate the dependence of $\tau_{C}$ on quasar parameters we fit the data\footnote{Here we treat masses and luminosities as independent quantities, although in practice both are based, to some extent, on the observed monochromatic luminosities.} with a function of the form:
\begin{equation} \label{eq:timescale}
\log \tau_{C} = a_{0} +a_{1} \log(1+z) + a_{2}(\log L-45) + a_{3} (\log M-8).
\end{equation}
The results of these fits are shown in Table  \ref{tab:FitTo005}. Noticing that the returned dependence on redshift is negligible, we also fit the same functional form with the redshift term forced to be zero, and find no significant change to the other two terms. These fits confirm that the main dependence of variability is with luminosity, with a much weaker dependence on black hole mass. We also note here that although the details of the fit will change with the choice of  amplitude of variability, C, the exact choice of C is inconsequential for our main conclusions (the clear dependence with L, little to no dependence with mass or redshift).\\

\begin{deluxetable}{llllr}  
\tablecolumns{10}
\tablewidth{0pt} 
\tablecaption{Fit to the time-scale of variability, Equation (\ref{eq:timescale}) }
\tablehead{   
  \colhead{$a_{0}$} &
  \colhead{$a_{1}$} &
  \colhead{$a_{2}$} &
  \colhead{$a_{3}$}&
  \colhead{$\chi^{2}/\mbox{d.o.f}$}  
}
\startdata
1.46 $\pm$ 0.02    & 0.02 $\pm$ 0.16 & 0.41 $\pm$ 0.04 & -0.09$\pm$0.03  & 2.93 \\  
1.46 $\pm$ 0.02  & \hspace{0.5 cm}$\equiv$ 0 & 0.41 $\pm$ 0.04 & -0.09$\pm$0.03 & 2.89  
\enddata
 \label{tab:FitTo005}
\end{deluxetable}

\begin{figure}[htp]
    \centering
    \includegraphics[width=.49\textwidth]{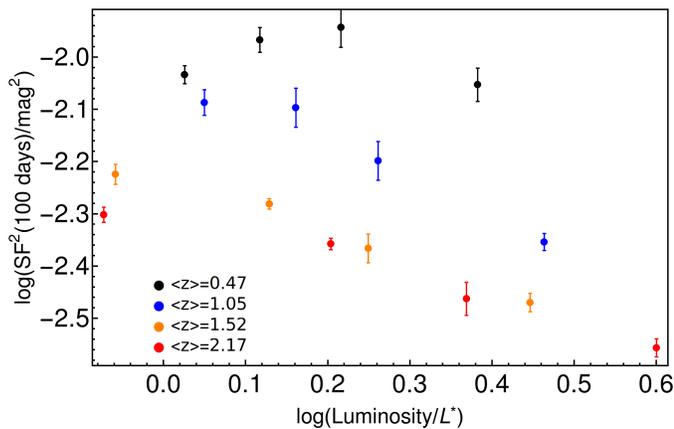}
    \caption{The dependence of the SF$^{2}$, measured at 100 days in the rest-frame, as a function of the luminosity relative to the knee of the quasar luminosity function at that redshift, $L/L^{*}$. While the anti-correlation with luminosity at a given redshift is still obvious, there is no longer the independence with redshift in this $L/L^{*}$ plot compared with that for simply L as in Figure \ref{fig:LPlot}.}
     \label{fig:Lstar}
\end{figure}

The lack of redshift evolution in the link between variability (whether in terms of amplitude or timescale) with luminosity is interesting.   The quasar luminosity function can be represented by a double power-law with a characteristic break luminosity $L^{*}$.  This break luminosity increases with redshift roughly as $L^{*} \propto (1+z)^{3-4}$ up to $z \approx$ 2 and then plateaus or even decreases at higher redshifts (e.g. \citealp{Hop07}, \citealp{Ued14}). In a simple phenomenological model, \cite{Cap15} suggested that this $L^{*}$ arose from the combination of a characteristic Eddington ratio, $\lambda^{*}$ , the characteristic Schechter mass of the host galaxies, $M^{*}$, and the $m_{bh}/m_{star}$ ratio. It was suggested that the $(1+z)^4$ evolution in $L^{*}$ was produced by a roughly equal $(1+z)^2$ evolution in $\lambda*$ and $m_{bh}/m_{star}$ (see \cite{Cap15} for details).  In this model, the luminosity of an AGN relative to the $L^{*}$ of the population (for $L > L^{*}$) emerges as a useful diagnostic of luminous AGN.  The above fits make clear however that quasar variability seems to be linked to $L$ and not to $L/L^{*}$.  This is illustrated graphically in Figure \ref{fig:Lstar}. We see that if the amplitude of variability is plotted against $L/L^{*}$, there is no agreement across different redshifts bins, as expected as we do not see any significant redshift evolution in the SF (see Table~\ref{tab:FitTo005}), while $L^{*}$ is known to strongly evolve with redshift (as noted above). \\

It is intriguing that the timescale of variability is most strongly correlated with luminosity, with the scaling $\tau \propto L^{0.4}$, rather than black hole mass.  This can be compared with the inferred time-scale of variability in the thin-disk accretion model. For an accretion disk around a black hole with a mass $M$ and a given luminosity $L$ the temperature of the disk is given by \citep{Net13}
\begin{equation}
T \propto M^{-1/2} L^{1/4} \left(\frac{r}{r_{g}} \right) ^{-3/4},
\end{equation}
where the $r_{g}$ is the gravitational radius of the black hole, $r_{g} \propto M$. This expression ignores the spin term.

From this we can follow \cite{Mac10} to derive a characteristic location within the accretion disk that is connected with a certain temperature, and thus with a corresponding emission wavelength: 
\begin{equation}
r_{\lambda} \propto M^{1/3} L^{1/3} \lambda^{4/3}.   
\end{equation}
From Kepler's law we have that 
\begin{equation}
t_{\rm{dyn}} \propto r^{3/2} M^{-1/2}.
\end{equation}
Combining these two expressions and identifying $t_{\rm dyn}$ as the time-scale of variability, we could expect that the timescale of variability at a single wavelength should depend on luminosity alone as
\begin{equation}
\tau \propto L^{1/2}.
\end{equation}
This is not dissimilar to the fitted $\tau \propto L^{0.4}M^{-0.1}$ behaviour we find in (i)PTF data (see Table \ref{tab:FitTo005}).   Obviously this is the simplest possible model.  In reality, the observed flux at given wavelength is produced from a wide range of radii and therefore mixing various time-scales. Such kind of ``mixing'' could significantly weaken any underlying luminosity dependencies, making them harder to identify in our data.

We note that this line of argument was already explored in \cite{Mac10}, who however noted that the de-correlation timescales in their damped random walk model did {\it not} appear to correlate with luminosity in this simple way.  As noted above, we prefer not to ascribe a physical significance to the (possible) breaks in the SF, because of the observational artefacts that can easily produce them.    Rather, we instead point out that the simple luminosity-only dependence is observed, with approximately the correct index, if we interpret the relevant ``timescale'' to be the time taken to reach a given level of variability.   One caveat is that while the luminosities of the quasars used in the above analysis are nominally bolometric luminosities, they are in fact ultimately derived from, and are proportional to, the monochromatic luminosities $L_\lambda$ at 5100, 3000 or 1350 \AA. Consideration of $L_\lambda$ in the context of a thin accretion disk would modify the expected relation and introduce a mass dependence of the form

\begin{equation}
\tau \propto L_\lambda^{3/4}  M^{-1/2}.
\end{equation}

\section{Variation of the PSD slope} \label{sec:Slope}

In this part of the paper we will investigate the slopes of the PSD of objects, which may contain information on the process driving the variability.   We model the PSD with a simple power-law $PSD \propto f^{-\alpha}$.  Different values of $\alpha$ will correspond to different correlation properties of the driving process.  For instance, $\alpha=0$ corresponds to the white noise of a completely random process without memory, while $\alpha=2$ is indicative of a random walk process.  Since the PSD and SF are closely connected quantities, we would expect the effects visible in a PSD analysis to also be seen in the SF analysis described above, although possibly modified due to the different types of systematics affecting each type of analysis.  One difference is that the PSD analysis is more readily carried out on an object by object basis whereas the SF is ideally suited to combining observations of objects within a sample. 
In general, for a PSD described with a power law $PSD (f) = \kappa f^{-\alpha}$ and $1<\alpha<3$ the SF$^{2}$  will have the following analytical form (e.g. \citealp{Emm10}):
\begin{equation}\label{eq:SFFromPSD}
\mbox{SF}^{2} (\Delta t) = -2^{\alpha} \kappa \pi^{\alpha-1} \Gamma(1-\alpha) \sin \left( \frac{\alpha \pi}{2} \right) \Delta t^{\alpha-1}. 
\end{equation}
where $\Delta t = f^{-1}$. From this we can see that a PSD with slope of $\alpha$ should correspond to a SF$^{2}$ with slope of $ \alpha-1$.

\subsection{Slope of PSDs in the PTF sample}

\begin{figure*}[htp] 
    \centering
    \includegraphics[width=.99\textwidth]{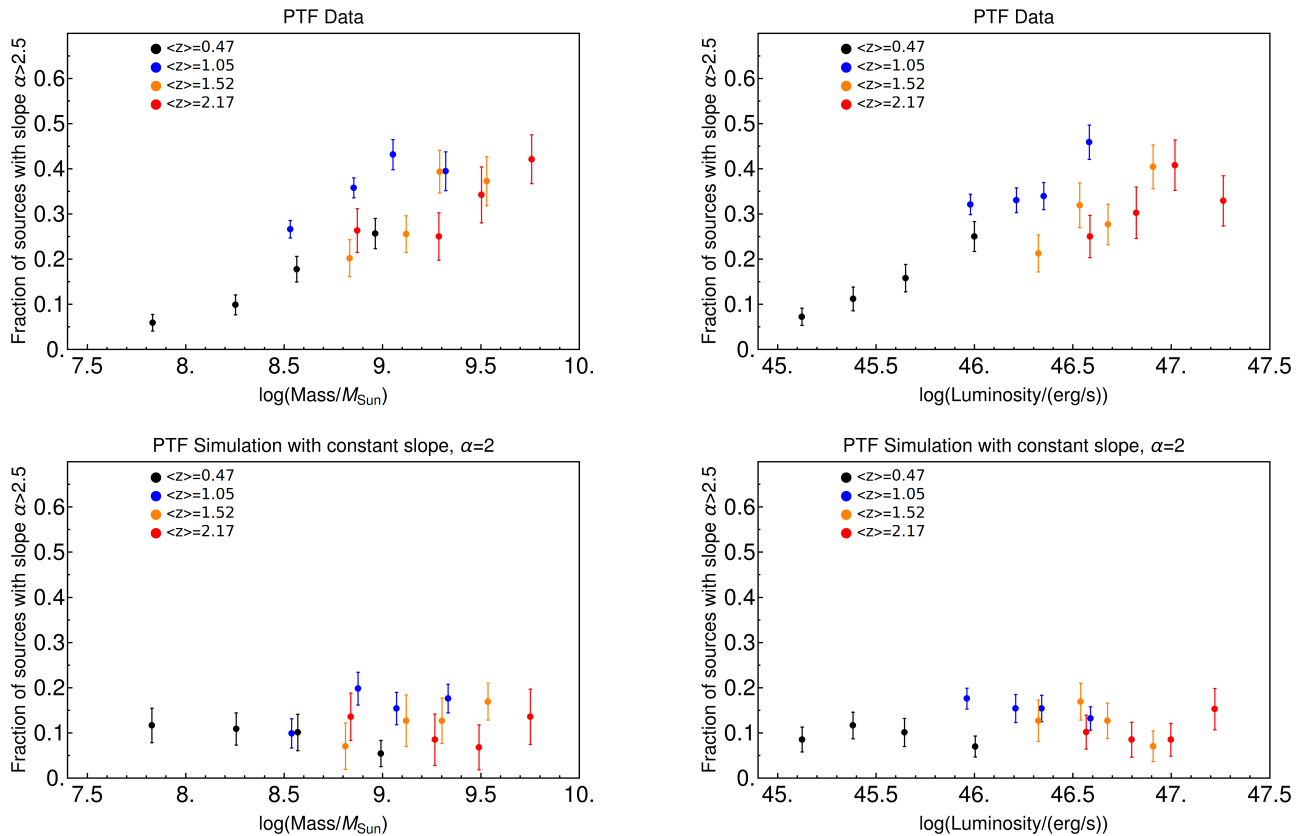}
    \caption{The dependence of the PSD slope, $\alpha$, with mass (left) and luminosity (right). In each panel, the colours denote different redshift bins.  We show the fraction of sources for which the CARMA modelling gives a slope $\alpha >$ 2.5.  In the upper panels we show the results from the (i)PTF data, while in the lower panels we show the results from the simulation in which artificial sources with a fixed input slope of $\alpha=2$ where observed with the cadence and photometric uncertainties of the actual data.  Whereas the real data shows a systematic increase with luminosity and/or black hole mass, the artificial data shows no such effect, suggesting that we are observing a real effect of steepening of the PSD slope with luminosity and mass in the quasars.  }
    \label{fig:PSDMain}
\end{figure*}

Fitting the PSDs of the 2,200 available objects for which the PSD can be constructed (see Section \ref{sec:DecTestPSD} above) yields a range of slopes between $1.5 < \alpha < 4$.   In Figure \ref{fig:PSDMain} we show the fraction of quasars with slopes steeper than $\alpha = 2.5$, choosing this threshold to be significantly above the random walk value of $\alpha = 2.0$.   The whole sample was first split in 4 redshift bins with the same procedure as described in Section~\ref{ConSF}, so that each redshift bin contains the same number of objects. We show results for two ways in which we further split the sample. 
In the case shown on the left hand side of Figure \ref{fig:PSDMain}, we have further divided each redshift subsample into 4 luminosity bins.
Alternatively, in the right hand side of Fig.~\ref{fig:PSDMain} we have instead divided each redshift subsample into 4 mass bins. 
We find this binning approach to be preferable, at this stage, to creating the larger number of 64 bins in mass-luminosity space, as we did in the structure function analysis, simply due to our desire to maintain statistical robustness given the much smaller number of objects for which the PSD can be determined. 
We will explicitly discuss the mass/luminosity dependence below. We have also discarded objects whose PSD appear to be inconsistent with the single power law description, i.e. which show strong bending. Specifically, we discard any object for which the best fit single power-law fit deviates beyond the 95\% confidence regions of the PSD analysis. This removes  20\% of the quasars. We also show in the lower two panels of Figure \ref{fig:PSDMain} the results that would be obtained from the observational uncertainties if all sources were in reality described by the $\alpha=2.0$ PSD of a pure random walk. These simulations were prepared as described in Section \ref{sec:DecTestPSD}. After the simulated data was created, exactly the same analysis procedure was applied to it as to the real data.\\

There is a clear trend in Figure \ref{fig:PSDMain} for quasars with higher luminosity and/or black hole mass to have, on average, steeper PSD.  The simulations with uniform $\alpha = 2$ do not show this steeping effect, and the fraction of (apparently) steep PSDs among simulated light curves stays at roughly 10\% independent of luminosity or mass. While the lowest luminosity/mass systems are consistent with this fraction, i.e. with all sources having $\alpha = 2$ in reality (given the observational uncertainties), the quasars at higher luminosity/mass have systematically ``redder" PSDs.  The effect can be seen across redshifts, but it is also visible in individual redshift bins, making clear that it is not a bias caused by the shifting time-dilated observing window.  This is further verified as the effect is still visible when we construct samples that have exactly the same span of rest-frame time.\\

This effect should be visible in the larger ensemble SF analysis described above.  In Figure \ref{fig:SFSlopes} we show the slope $g_{1}$ of the SF, i.e. the slope of the short timescale variability (see Equation \ref{eq:SFeq}), using the same binning and symbols as for the PSD analysis in Figure \ref{fig:PSDMain}.  The grey dashed lines show the SF$^{2}$ slope that would be expected from the PSD analysis using Equation \eqref{eq:SFFromPSD}. When doing this we have for simplicity assumed that all of the quasars in a single bin have exactly the same PSD dependence, defined by the median $\alpha$ in that bin. This is the simplest assumption one can make and is used here to check the consistency of the results and should not be interpreted as a precise physical statement.  \\

We see that the results are quite consistent. The same dependence of slope on luminosity and mass is seen in the SF$^{2}$, with the more luminous/massive systems having steeper slopes on average. The results from the SF$^{2}$ analysis alone are however not conclusive, and the dashed line offers only a 25\% improvement when measuring $\chi^{2}/\mbox{d.o.f.}$ compared to the best horizontal of constant slope, ($\chi^{2}/\mbox{d.o.f.} \mbox{ (constant)}=3.72$, against $\chi^{2}/\mbox{d.o.f.} \mbox{ (dashed line)}=2.82$). However, we find it encouraging that both types of analysis show consistent results.  We again emphasize that the dashed line in Figure \ref{fig:SFSlopes} is not the fit to the data - it is an \textit{expectation} derived from the PSD analysis.\\

\begin{figure*}[!btp] 
    \centering
    \includegraphics[width=.99\textwidth]{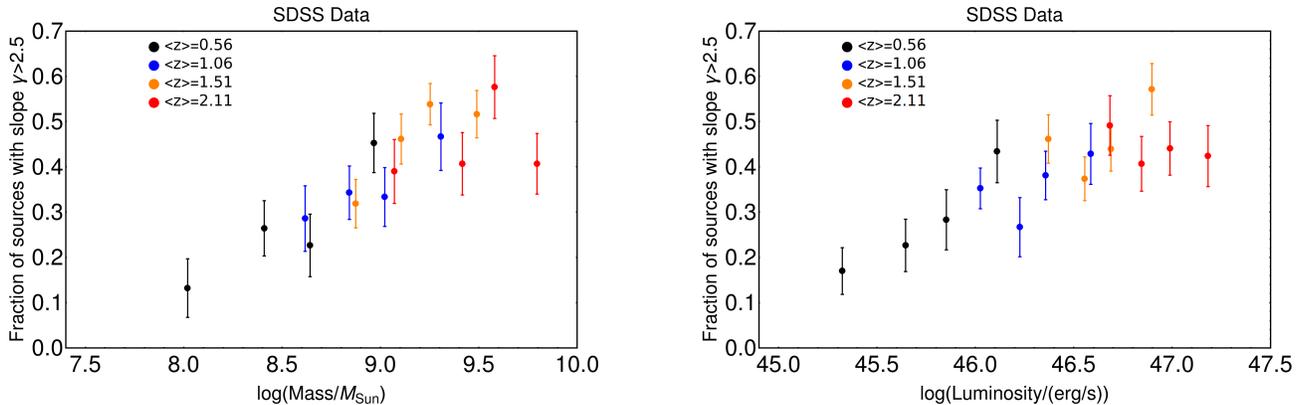}
    \caption{This plot is the same as the upper panels of Figure \ref{fig:PSDMain}, but with the analysis done for the SDSS Stripe 82 data. The same effect, with more massive and more luminous quasars displaying steeper slopes of their PSD, is observed.   }
    \label{fig:SDSSPSD}
\end{figure*}

\begin{figure}[htp] 
    \centering
    \includegraphics[width=.49\textwidth]{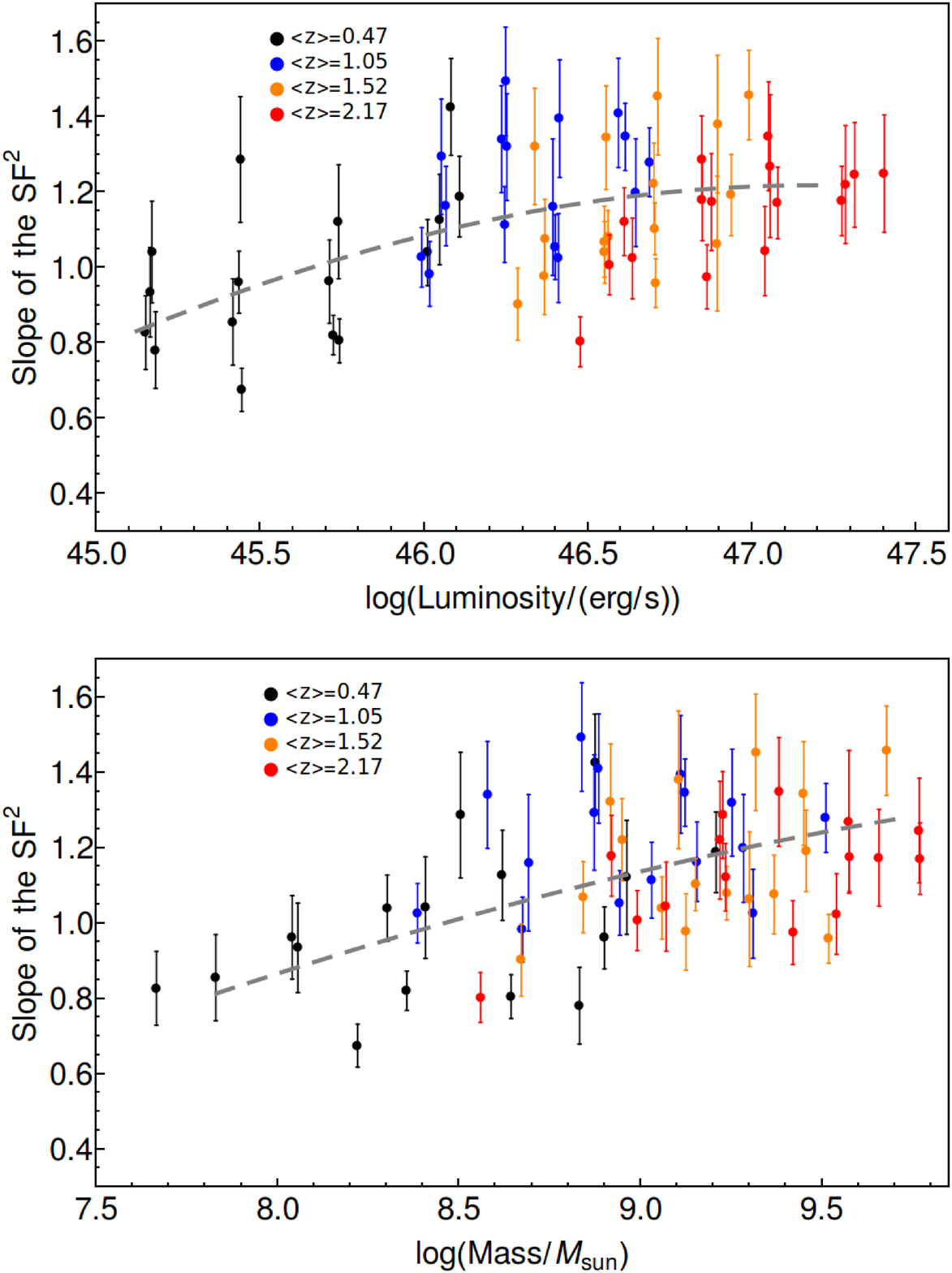}
    \caption{The short time-scale slopes of the structure function SF$^{2}$, i.e. $g_{1}$, as a function of luminosity and mass. The dashed line we show the \textit{expected} dependence from the PSD results. We see that the SF$^{2}$ results are consistent , albeit with greater noise, supporting the reality of this effect.}
    \label{fig:SFSlopes}
\end{figure}

As an additional check on the reality of this result, we perform the same PSD analysis with the $r$ band observations in the sample of SDSS Stripe 82 quasars. We perform exactly the same procedure as for our (i)PTF sample, selecting only well sampled light curves and splitting into the four redshift and then four mass or luminosity bins. The results are shown in Figure \ref{fig:SDSSPSD}. We see that we recover exactly the same trends as in our (i)PTF analysis, showing the strong dependence and the steeping of the slopes with luminosity and mass.   We conclude that this is likely to be a real effect. \\ 

To investigate further the connection between the steepness of the slope and mass and luminosity, we fit the dependence of the slope with a functional form
\begin{equation} \label{eq:SlopeFit}
g_{1} = b_{0} + b_{1}(\log L-45) + b_{2} (\log M-8).
\end{equation}
Given the weak effects of redshift in Figures \ref{fig:PSDMain}, \ref{fig:SDSSPSD} and \ref{fig:SFSlopes}, we do not include redshift as one of the parameters in order to reduce the degeneracy in our fits. In Table \ref{tab:FitTSlope} we show our results for different binning and analysis techniques.  We have first fitted the dependence of the SF$^{2}$ slopes, as derived in Section \ref{ConSF}, using the full sample split in redshift, luminosity and mass. Second, we have suppressed the information about the redshift and created new bins which were split only in luminosity and mass and for which we fitted the SF$^{2}$ with the same procedure. Third, we fit the results from the PSD analysis using 64 bins of redshift, luminosity and mass. We report the inferred values for the SF$^{2}$ slope by using the Equation (\ref{eq:SFFromPSD}), i.e. by assuming that PSD slopes and the SF$^{2}$ slopes differ by one. This is represented in the table with the (+1) symbol at the relevant position. Finally,  case (4) shows our fit to the results from our PSD analysis of the SDSS Stripe 82 sample using the same 64-bin approach. \\

\begin{deluxetable}{llllr}  
\tablecolumns{25}
\tablewidth{0pt} 
\tablecaption{Dependence of the SF$^{2}$ slope, fit with Equation \eqref{eq:SlopeFit} }
\tablehead{   
 \colhead{case$^{1}$} &
  \colhead{$b_{0}$} &
  \colhead{$b_{1}$} &
  \colhead{$b_{2}$} &
  \colhead{$\chi^{2}/\mbox{d.o.f}$}  
}
\startdata
1 & 0.82 $\pm$ 0.05    & 0.05 $\pm$ 0.06 & 0.13 $\pm$ 0.07 &  2.72 \\  
2 & 0.93 $\pm$ 0.05    & -0.02 $\pm$ 0.06 & 0.20 $\pm$ 0.06 &  3.10 \\
3 & 0.88 (+1) $\pm$ 0.05   & 0.04 $\pm$ 0.06 & 0.19 $\pm$ 0.06 &  1.71 \\ 
4 & 0.87 (+1) $\pm$ 0.06   & 0.0 $\pm$ 0.08 & 0.33 $\pm$ 0.09 &  1.50
\enddata
\tablenotetext{1}{Procedure of how each sample was prepared and which methods were used is described in the text}
 \label{tab:FitTSlope}
\end{deluxetable}

All of these results point towards a stronger dependence on the mass of the black hole than on the quasar luminosity. 
The fact that all these different types of analysis lead to similar conclusions is reassuring given the uncertainties and the different systematics affecting each method. \\

The existence of the strong degeneracy between the mass and luminosity in quasar samples means that, even though the dependence appears preferentially to be with mass, there will inevitably be an apparent secondary correlation with luminosity (and thus Eddington ratio). The uncertainties are large enough that a primary dependency on luminosity is only excluded at the $\sim$2 $\sigma$ level, depending on the type of analysis used. We also note that if the data taken from two lowest luminosity bins are removed the confidence contours enlarge enough that even pure luminosity dependence is allowed within 1 $\sigma$ level when using  the SF data shown in Figure 15 (described as Case 1 in the text above).  \\

The variation of the PSD and SF slopes has previously been hinted at in the recent Pan-Starrs study of \cite{Sim16}. In that study PSDs were constructed for a sample of X-ray selected AGNs in the COSMOS field. That work showed differences in the high frequency slopes of the PSD as a function of mass, although the small number of objects in the sample and the uncertainties associated with the measurement and the PSD method prohibited any strong conclusions. The PSDs were modelled as a broken power law, with two slopes.  As discussed above, we prefer to fit a single power-law to the PSD.   Clearly the much larger number of sources and greater time sampling of the (i)PTF observations relative to the Pan-Starrs coverage of COSMOS makes this effect much clearer.\\

Less directly, \cite{Koz16} recently conducted a re-analysis of the SDSS Stripe 82 data using the SF approach. When using the full sample without any cuts (in redshift, mass or luminosity), this study tentatively found a somewhat steeper slope of the SF ($g_{1}=1.1 \pm 0.16$) than would be expected from the pure random walk.  This is also consistent with our results since only in the lowest luminosity/mass bins are the sources consistent with all having random walk PSDs (see Figures \ref{fig:PSDMain}, \ref{fig:SDSSPSD} and \ref{fig:SFSlopes}). More interestingly, that study also reported a weak correlation of the SF slope and the luminosity of the AGNs, ranging from $g_{1} \approx 0.9$ for the lowest luminosity and lowest redshift objects  ($\log$ (L/erg s$^{-1}$) $\sim 45$), reaching $g_{1} \approx 1.3-1.4$ for the high luminosity objects in the sample ($\log$ (L/erg s$^{-1}$) $\sim 47.5$). 
We also note in passing that the effect can in fact be seen, although it was not extensively discussed, in the early analysis of the SDSS Stripe 82 data by \cite{Voe11}, as a difference in the SFs of high and low mass quasars in his Figure 8.\\

The fact that this effect is clearly seen using two different analysis techniques in our very large (i)PTF data set, and is also present at lower significance in other data sets using both the PSD and SF approaches (i.e. the PSD method in PanStarrs \cite{Sim16}, the SF method in SDSS - Stripe 82 \cite{Koz16}) makes us confident that this subtle effect is nevertheless real.  \\

\section{Discussion}

%
%
In the preceding section we have pointed out several interesting results linking quasar variability with their physical properties, using both the SF and PSD formalisms to describe the variability. Using the SF formalism we were able to show that the amplitude of variability, defined as the value of SF$^{2}$ at a particular rest-frame timescale (100 days), or, largely equivalently, the timescale to reach a certain value (0.005 mag$^2$), is most strongly correlated with the luminosity of the quasar, with little or no dependence with with we black hole mass or redshift. When using the PSD approach, we find clear evidence for variations of the PSD slope, with quasars powered by the higher mass black holes exhibiting steeper PSDs. We also find that the slopes from the SF analysis are also consistent with this steepening trend. The effect is seen when splitting the sample in either luminosity or mass bins, but a combined analysis tentatively suggests that mass is the driver of the effect, although luminosity (or Eddington ratio) can not be conclusively excluded.\\

This work, as well as several others (\citealp{Mus11}, \citealp{Ede14}, \citealp{Koz16}) have highlighted observational shortcomings in the damped random walk model of quasar variability, in which a random walk at short timescales flattens at longer timescales in the damped part.   In particular, there is now rather good evidence that a non-negligible fraction of quasars exhibit variability which have significantly redder PDSs at low frequencies than expected from the random walk, with this fraction depending on the black hole mass.\\

Could this behaviour reflect a situation in which quasars display damped random walk behaviour at medium and long time scales and a steeper spectrum at short time scales? The time scale at which the model switches from the "steep" to the random walk behaviour would need to be connected with the mass of the black hole, in order to reproduce the mass dependence of the slope that we see in the data. This time scale would need to be at quite short time-scales, below our detection threshold ($\sim$ 0.1-10 days) for the low mass AGNs where we do not observe any steepening, and much longer ($\sim $10-1000 days) for massive systems where we observe steeper slopes. For instance, in low mass ($10^{7}$ $M_{\odot}$) AGN, \mbox{Zw 229-15}, which was monitored by Kepler, the time-scale of switching from steep slope to the random walk slope is at $\sim$ 5 days and as such the steep part of the PSD would not be observed in our data \citep{Ede14}.   \\

In this scenario it might be thought that the shift in timescale to achieve the steepening was associated with the shift in timescale that we have argued is equivalent to the change in variability amplitude.  It is not easy to make such a model work in practice, the change in slope being effectively too large.  This suggests that two separate processes may be involved in the change in amplitude/timescale, and the spectra steepening, as also implied by the apparently different dependences on luminosity and mass.

\section{Conclusions}
The main aim of this paper has been to characterise the optical variability of a very large number of quasars in the Palomar Transient Factory  (PTF) and intermediate Palomar Transient Factory  (iPTF) surveys. We have analysed light curves from 28,096 AGN with the total of 2.4 million photometric data-points. All of these light-curves were re-calibrated specifically for this work. The quality of the (re-)calibration is very high with the vast majority of the AGN sample not showing any excess variability at the shortest scales. We have used both the structure function (SF) and power spectral density (PSD) formalism to characterize the variability of the quasars and to search for connections of the variability with redshift, black hole mass and luminosity. Our main conclusions can be summarized as follows:

\begin{itemize}
\item The amplitude of variability, defined in this work as the value of the SF$^{2}$ at a rest-frame time scale of 100 days, exhibits a clear anti-correlation with luminosity and little or no variation with mass or redshift. We find that simple epoch-independent anti-correlation with luminosity is clearly preferred over connections of variability with the normalised luminosity $L/L^*$, where $L^*$ is the break in the quasar luminosity function.  

\item  The time scale $\tau$ at which the variability reaches a given amplitude, which in this work we set at SF$^{2}\left(\tau\right)=0.005\,{\rm mag}^{2}$, is closely connected to the variability amplitude described above. We again find that $\tau \propto L^{0.4}\,M^{-0.1}$, without any redshift dependence. 
We note that this trend is broadly consistent with the expectation from a simple Keplerian accretion disk model, where the orbital time scale follows $\tau_{\rm dyn} \propto L^{0.5}$, with no mass dependence.

\item There is a clear variation of the slopes of the PSDs, with many quasars that are steeper than what is expected from a random walk model. 
Quasars with high mass and/or luminosity tend to have steeper PSD slopes. 
The steepening of the PSD slopes is consistent with the observed slopes of the SF functions, and substantially strengthens the indications of this effect in other recent variability studies. 
There is some evidence in our own data that the dependence is primarily with the black hole mass rather than the luminosity. 

\item
The observed dependence of the PSD slope with the black hole mass or AGN luminosity could be reproduced in a model in which PSD exhibits steeper slopes below a certain time-scale, with this time-scale being dependent on mass/luminosity, but this would require a strong compensating effect on the amplitude. 

\end{itemize}

We are very grateful to Shri Kulkarni and the PTF and iPTF collaborations for their generous granting of access to their photometric database. We thank the referee for their useful comments. This work has been supported by the Swiss National Science Foundation.

\bibliography{Paper001}

\renewcommand{\theequation}{A-\arabic{equation}}
  
  \setcounter{equation}{0}    
  
\section*{Appendix A: Completeness and depth of the survey} \label{sec:ComDep}

\begin{figure}[h!]
    \centering
   \includegraphics[width=.49\textwidth]{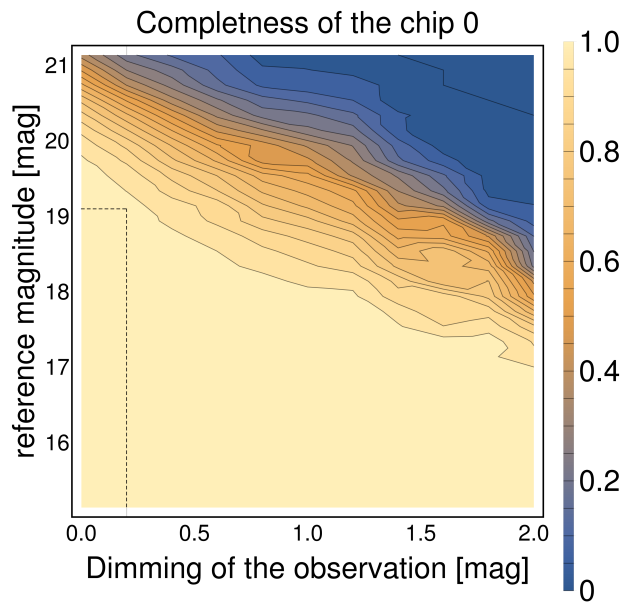}
    \caption{The completeness of the survey, as in the single representative Chip 0, as a function of the observed magnitude of the source and the dimming of the observation (see text for details). Each contour denotes 5\% change in the probability that a given object is included in the SExtractor sample. The dashed lines show the region from which we select the observations to be used in this work. }
     \label{fig:Compl}
\end{figure}

In this Appendix, we wish to characterize the completeness of the (i)PTF survey. In Figure \ref{fig:Compl} we show the completeness, i.e. the probability that an object is detected as a function of its ``reference magnitude'', as defined in the main body of the text and below, and the atmospheric absorption (mostly due to clouds) of the observation.  For simplicity, only sources in chip 0 are shown - all of the other chips display similar behaviour. As described in Section \ref{sec:Cal}, we assign to each object its reference magnitude as a mean value of the measurements from the 5 brightest observations, while the dimming of the observation is deduced as the median difference of the observed brightness of the reference objects to their reference values in that particular observation. As expected, the completeness stays roughly constant along the lines of constant observed brightness, i.e. along the line defined by subtracting the dimming of the observation from the reference magnitude. The dashed line delineates the region from which we select our AGN sample. We select only objects which are brighter than $r=19.1$ magnitude and we only take into account observations taken in the clear conditions, i.e. with atmospheric dimming below 0.2 magnitude. This conservative cut is applied in order to make sure that we are only minimally biased in our estimate of variability; if we included fainter objects in our selection we would not be able to observe them during their fainter phases, biassing the results.  We also see evidence that, when there is significant cloud absorption, the absorption varies significantly over the large field of view of the camera, as might be expected given the short exposure times. 

\section*{Appendix B: Comparison of calibration with Ofek et al. (2012)} \label{sec:Ofe}

\begin{figure}[h!] 
    \centering
   \includegraphics[width=.49\textwidth]{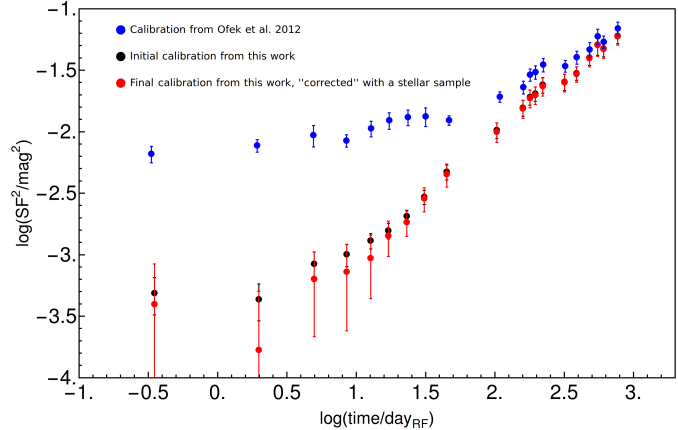}
    \caption{Comparison of the results for the quasar SF$^{2}$ constructed using the standard (i)PTF calibration described in \cite{Ofe12}, the re-calibration described in this work and the final correction using the stellar SF$^{2}$.  The re-calibration is successful in significantly reducing the systematics which can dominate the variability at short time scales. After correction by the stellar SF$^{2}$, the final quasar SF$^{2}$ is generally consistent with zero as $\Delta t$ approaches zero, as required.}
    \label{fig:CalibrationComparison}
\end{figure}

In this Appendix we will briefly compare the calibration used in this work and the original calibration of the (i)PTF survey presented in \cite{Ofe12}. As an example, in Figure  \ref{fig:CalibrationComparison} we show the initial AGN SF$^{2}$ for the sample constructed at redshift $\left\langle z \right\rangle =1.05$, luminosity of  $\left\langle \log L \right\rangle =46.04$ erg/s and mass $\left\langle \log M \right\rangle =8.72$ $M_{\odot}$. Note that, for ease of comparison, this is the same sample as considered in Appendix C. The residuals are large when using original calibration, while when using our improved methods we see significant reduction. The small excess at short times that can be seen also when using our calibration can be accounted for in statistical sense by correcting with stellar sample, as described in Appendix D. \\

\begin{figure}[h!] 
    \centering
   \includegraphics[width=.99\textwidth]{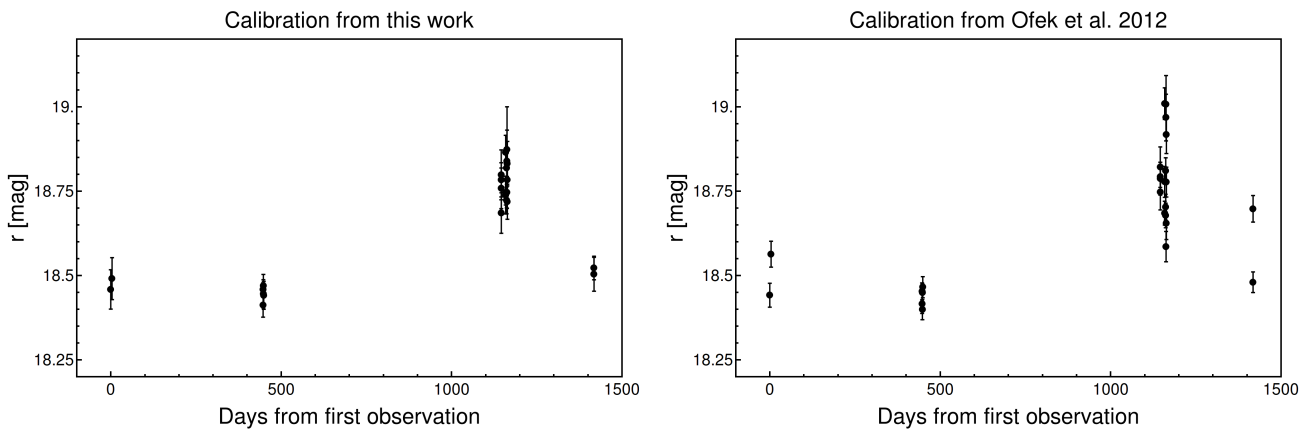}
    \caption{Light curve for the representative (randomly chosen) quasar SDSS J102255.21+172155.7. On the left hand side we show the light-curve using the re-calibration from this work and on the right hand side we show the light-curve with the original calibration from \cite{Ofe12}. The increased consistency of same-night observations is apparent. Notice also the smaller scatter of the group of the points at $\sim$ 1150 days. The calibration improvements have been critical in enabling us to measure the variability at short ($\sim$ 10 days) time scales.}
    \label{fig:SingleExampleCalibration}
\end{figure}

This difference can be attributed to two factors. Firstly, we find that our error estimates, which are taken from the spread of the calibrator objects around their means, are consistently around 30 \% larger then pure SExtractor error estimates used in \cite{Ofe12}, indicating that the SExtractor does not capture fully the measurements uncertainties in the survey. Additionally, we have significantly improved the stability of the survey; data points taken with small time separation are highly consistent with each other, which is often not the case when using the original calibration. Example of this is shown in Figure    \ref{fig:SingleExampleCalibration}.

\section*{Appendix C: Colour dependence of the calibration} \label{sec:Col}

Given that there were no explicit color terms in the calibration (see Section \ref{sec:Cal}) we investigate if there are significant color-dependant trends in the final calibration product used. In order to do this we construct the SF$^{2}$ for the  sample of stars for which the color is known from the SDSS. We select 28667 available reference objects (stars) from 100 randomly selected fields (for conditions to be selected see Section \ref{sec:Cal}). We calibrated these objects with the same procedure with which we calibrated AGNs. We then split our sample into 20 bins given their u-g color and created ensemble SF$^{2}$ functions for each sub-sample, using the method described in Section \ref{sec:SF}. Finally we calculate the "Mean SF$^{2}$" as:
\begin{equation}
\left\langle  SF^{2} \right\rangle    = \frac{1}{N} \sum^{N}_{i} SF^{2} (t_{i})
\end{equation}
where we averaged over all $N=20$ time bins in which we estimated SF$^{2}$.  Errors are indicating the spread of the SF$^{2} (t_{i})$ measurements. In an ideal case, and if there was no underlying stellar variability, all of the points would be at exactly zero, indicating that we have perfectly captured and subtracted errors in the sample.  We see that the values of $\left\langle  SF^{2} \right\rangle$ are slightly positive indicating that there are some residuals in our SF$^{2}$ estimation, but these do not depend on colour, suggesting that stellar variability effects are small.  In general our estimated errors account for around 90\% observed variance. Our procedure for dealing with this problem when constructing AGN SF$^{2}$ is further elaborated in Appendix D.\\
	
\begin{figure}[h!]
    \centering
   \includegraphics[width=.49\textwidth]{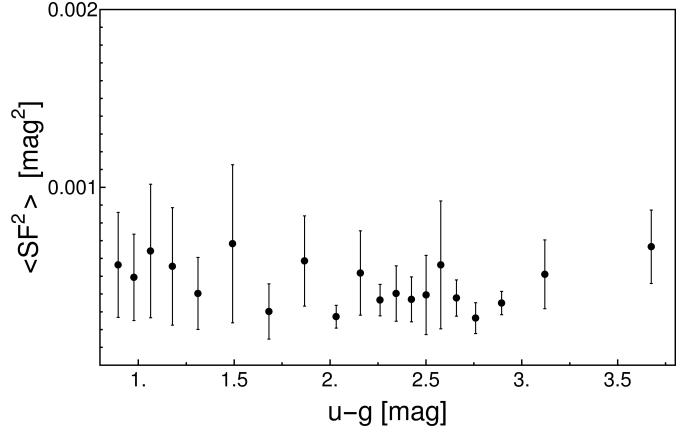}
    \caption{The SF$^{2}$ for stars (averaged over all $\Delta t$) as a function of their colour, as constructed using the procedure described in the text. There is no significant color-dependence, suggesting (a) that instrumental effects are likely not colour-dependent and (b) that the stellar SF$^2$ is probably not dominated by intrinsic stellar variability but rather by uncontrolled systematic uncertainties in the photometry that presumably would also be present in the quasar data.  It should be noted that strongly varying sources, with Stetson variability parameter $J \gt 4$. were removed from the staple sample - see text for details.}
     \label{fig:ColDep}
\end{figure}

\section*{Appendix D: Reduction of SF$^{2}$ residuals using the stellar sample} \label{sec:sf2res}

As explained in Section \ref{sec:Col}, our estimated errors on the measurements during (re-)calibration account for around $\sim$ 90\% of the observed variance, causing our estimated SF$^{2}$ to be slightly above zero. In order to account for this when analysing the AGN SF$^{2}$ we construct an equivalent sample of stars which are matched to the AGNs in terms of their brightness. We construct the SF$^{2}$ for these stars using the same procedure as for the AGNs. To take into account that some stars are also variable we do the analysis for the full sample of stars and also for the sample of stars for which we are sure that are non-variable, i.e. that have Stetson $J$ index smaller than 4. The Stetson J-index is defined as (\citealp{Ste96}, \citealp{PrW14})
\begin{equation}
J=\sum^{N-1}_{i} \mbox{sign}(\delta_{i}\delta_{i+1}) \sqrt{|\delta_{i}\delta_{i+1}|} \hspace{1 cm} \mbox{ where }  \hspace{0.5 cm}\delta_{i}= \sqrt{\frac{N}{N-1}} \frac{x_{i}-\mu}{\sigma_{i}} 
\end{equation} 
where $x_{i}$ are measurements values, $\mu$ mean value and $\sigma_{i}$ are measurements errors. Stetson $J$ tends to 0 for non-variable stars and is large when there are adjacent measurements are discrepant. We then construct the final SF$^{2}$ as the mean of these two samples in each time bin and conservatively assume largest possible errors from these two samples.\\

\begin{figure}[h!] 
    \centering
   \includegraphics[width=.99\textwidth]{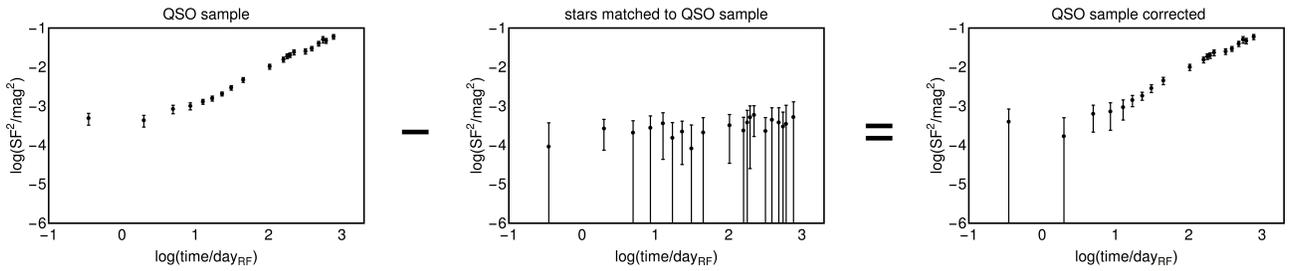}
    \caption{Schematic representation of our procedure to create the final SF$^{2}$ estimate for a quasar sample. Starting from the initial SF$^{2}$ (shown on the left) we  subtract the stellar SF$^{2}$ constructed from stars of similar magnitude, to reach the final, corrected quasar SF$^{2}$ shown on the right.  It should be noted that strongly varying sources, with Stetson variability parameter $J \gt 4$, are treated as described in the text.  }
    \label{fig:SFConstruct}
\end{figure}

When constructing the final AGN SF$^{2}$ for the quasars we subtract the SF$^{2}$ observed for the stars to get a final estimate of the AGN SF$^{2}$, which we then use in the analysis described in the main body of the manuscript. The errors in the final AGN  SF$^{2}$ are derived by adding in quadrature the errors from the initial AGN SF$^{2}$ and from the stellar sample SF$^{2}$.  The procedure is sketched in Figure \ref{fig:SFConstruct} for the sample constructed at redshift $\left\langle z \right\rangle =1.05$, luminosity of  $\left\langle \log(L/\mbox{erg s}^{-1}) \right\rangle =46.04$  and mass $\left\langle \log M/\mbox{M}_{\odot} \right\rangle =8.72$. 
Using this procedure we are able to fully capture the uncertainty in our estimates of SF$^{2}$ for a large number of AGN SF$^{2}$ in a sense that our final estimates are consistent with zero on the shortest time scales probed.

\end{document}